\documentclass[12pt,preprint]{aastex}
\newcommand{\kms}{{\,\rm km\,s}^{-1}} 
\newcommand{\ksm}{{\,\rm km}\ {\rm~s}^{-1}\ {\rm~Mpc}^{-1}}

\def\la{\mathrel{\hbox{\rlap{\hbox{\lower4pt\hbox{$\sim$}}}\hbox{$<$}}}}
\def\sun{\hbox{$\odot$}}
\renewcommand{\mag}{\mbox{$\;$mag}}

\begin{document}
\title{COMPARISON OF DISTANCES FROM RR LYRAE STARS, THE TIP
    OF THE RED-GIANT BRANCH AND CLASSICAL CEPHEIDS} 
\author{G. A. Tammann}
\affil{Astronomisches Institut der Universit\"at Basel,\\
       Venusstrasse 7, CH-4102 Binningen, Switzerland}
\email{g-a.tammann@unibas.ch}
\author{Allan Sandage}
\affil{Observatories of the Carnegie Institution of Washington,\\
       813 Santa Barbara Street, Pasadena, CA 91101}
\author{B. Reindl}
\affil{Astronomisches Institut der Universit\"at Basel,\\
       Venusstrasse 7, CH-4102 Binningen, Switzerland}
\email{reindl@astro.unibas.ch}

\begin{abstract}
The extragalactic distance scale relies heavily on Cepheids. However,
it has become clear from observations and pulsation models that the
slope and zero point of their P-L relations differ from galaxy to
galaxy. 
This makes the determination of Cepheid distances complex and calls
for an independent test of their differences. 
The test is provided by RR\,Lyrae star distances of 24 galaxies which
calibrate the tip of the red-giant branch 
(TRGB; $M_{I}^{\rm TRGB} = -4.05$), which in turn confirms the adopted
Cepheids distances on our 2006 distance scale in 18 cases to within
$0.1\mag$ on average. 
Relative  SN\,Ia and velocity distances deny a remaining significant 
metallicity effect of the adopted distances. The new support for these
Cepheid distances increases the weight of our previous calibration of
the SN\,Ia luminosity and of the 21cm line width - luminosity (TF) 
relation. The value of $H_{0} = 62.3 \;(\pm5)$ is confirmed on all
scales.  
\end{abstract}

\section{INTRODUCTION: BAADE'S EARLY ATTEMPT TO COMPARE RR LYRAE
          VARIABLES AND CLASSICAL CEPHEIDS IN M\,31}
\label{sec:01}
In the late 1940s and early 1950s Walter \citet{Baade:48} and
Edwin \citet{Hubble:51} had formulated their research plans for
observational cosmology using the new Palomar 200-inch telescope.
At various times in the 1930s Hubble had described his early
Cepheid distance scale to 
NGC\,6822 \citeyearpar{Hubble:25}, 
M\,33 \citeyearpar{Hubble:26}, and 
M\,31 \citeyearpar{Hubble:29}
in the local group as only ``reconnaissance'' studies. 
He had put the distance moduli of these galaxies all at about 
$\mu=(m - M) = 22.2$ ($D = 0.27\;$Mpc). Hubble then took the scale
outward using "brightest stars" (later shown to be HII regions) to the
M\,81/NGC\,2403 and M\,101 groups, both with 
$\mu = 24.0$ ($D = 0.63\;$Mpc) on his scale, 
and ultimately to the Virgo cluster where he adopted 
$\mu = 26.8$ ($D = 2.3\;$Mpc). 
Although he was fully aware of the time-scale difficulties given by
this distance scale that gave a redshift-distance ratio (the Hubble
constant) of about $530\kms$\,Mpc$^{-1}$ (the units are assumed in the
rest of this paper) with its small expansion age of 
$1/H_{0} = 1.8\;$Gyr, he nevertheless believed in this scale as late
as 1952. This can be seen in Holmberg's \citeyearpar{Holmberg:50} use
of Hubble's distances in his survey of the groups centered on M\,81,
M\,101, and the Virgo cluster, distances that had been recommended by
Hubble to him during his Mount Wilson stay to obtain observations
of galaxian magnitudes with the 60-inch telescope. 
It is also seen in Hubble's reluctance in 1952 to fully accept Baade's
revision by about $1.5\mag$ for the distance moduli of M\,31 and M\,33,
shown by the use of his scale in the study
\citep{Hubble:Sandage:53} of the bright blue super-giant variables in
each, with only a footnote to the new scale by Baade. (That footnote
was inserted into the manuscript by one of us who, at the time, was
Baade's student and Hubble's assistant. Hubble only gave his hesitant
approval to the footnote in the late drafts of that paper).

     Baade had been working in the 1930s on extending the
magnitude scales of \citet{Seares:etal:30} in the Mount Wilson
Catalogue of Selected Areas, particularly for SA 68, to the faintness
required by Hubble's Cepheids in the three local group galaxies of 
NGC\,6822, M\,31, and M\,33. By the time Baade had resolved the bulge
of M\,31, the face of NGC\,205, and the two dwarf companions of M\,31,
NGC\,147, and NGC\,185, into stars \citep{Baade:44a,Baade:44b} 
and had identified these stars as the tip of the red-giant branch in
globular clusters (often named the Baade Sheet in external galaxies
and now called the TRGB), he had corrected Hubble's magnitude scales
(which he often called Hubble's ``enthusiastic'' magnitudes), and
changed Hubble's M\,31 distance modulus \citep{Baade:44a} to be 
$\mu = 22.4$, which however was only $0.2\mag$ larger than Hubble's
value of 22.2. 

     By 1948 Baade had anticipated that he should have detected
RR\,Lyr stars in M\,31 with the 200-inch telescope starting near
$V = 22.2$ because he believed the mean absolute magnitude
of these variables was $M_{\rm pv} = -0.23$. This value had been
determined by \citet{Shapley:18} using the method of comparing the
apparent magnitudes of two distance-indicators (long period
Cepheids and RR\,Lyrae stars in this case) in aggregates of stars
(here the globular clusters) where they appear together, the
absolute magnitude calibration of one of which was taken to be
known. Shapley had used the method by assuming the Cepheids in
globular clusters were the same as the field Galactic Cepheids whose 
absolute magnitudes had been calibrated by \citet{Hertzsprung:13},
\citet{Russell:13}, and  \citet{Shapley:18} by the method of
statistical parallaxes.  
Although the theory of the comparison method is correct, the result
turned out to be wrong for the Galactic Cepheids but, remarkably,
almost correct for the RR\,Lyrae stars. It was 30 years later that
Baade made the distinction between the globular cluster Cepheids (of
his population II) and the population I Galactic Cepheids. It has
turned out that the statistical parallax calibration of the
Galactic Cepheids by Hertzsprung, Russell, and Shapley was too
faint by about 2 magnitudes, but Shapley's calibration
of the globular cluster Cepheids was close to what we know now to
be the correct calibration of the Population II Cepheids,
leaving Shapley's value of $\langle M_{V}\rangle = -0.23$ for the 
RR Lyrae stars reasonably correct for the purposes of the argument
made by \citet{Baade:54} at the Rome 1952 meeting (we adopt 
$\langle M_{V}({\rm RR\,evolved})\rangle = +0.52$ at [Fe/H]$ = -1.5$
in \S~\ref{sec:02}). 

     When Baade did not find the M\,31 RR\,Lyrae variables at 
$V = 22.2$ he had two choices. Either the absolute magnitude
calibration of the classical (population I) Cepheids was wrong, and
M\,31 was more distant than Hubble's modulus of 22.2, or Shapley's
calibration of the RR\,Lyrae variables near $M_{V} = 0$ was much too
bright. \citet{Behr:51} had suggested the error was in the Cepheid
calibration by nearly 2 magnitudes. Behr's paper was not cited by
Baade (although surely he knew of it because he was a voracious
devourer of the literature), probably because Baade believed his own
arguments to be decisive as he gave them at the 1952 Rome IAU meeting
rather than relying on the tricky method of statistical parallaxes of
Cepheids, sensitive as the result is to Galactic absorption.

     Hence, the method of comparing the apparent magnitudes of various
distance indicators with each other where they appear together began
with \citet{Shapley:18} in the globular clusters, and continued with
Baade's spectacular failure to find the M\,31 RR\,Lyraes at Hubble's
distance. The method was successful in the LMC and SMC with
the discovery by \citet{Thackery:54,Thackery:58} of RR\,Lyrae stars
near $V=19$ which he had also reported in the summary report of
Commission 28 at the 1952 Rome IAU meeting. 
It was Thackery's discovery of RR\,Lyrae stars in the LMC and SMC at
this faint magnitude, about two magnitudes fainter than predicted by
the then adopted zero point of the Cepheid P-L relation, that cemented
the truth of Baade's assertion that a change in the calibration of the
P-L relation was needed, and therefore that Hubble's distance to M\,31
must be too short. 

     The method of comparisons of Cepheids and RR\,Lyrae stars in
individual galaxies lay dormant until it was taken up again 
by \citet{Walker:Mack:88}, and was also greatly stimulated by the
arrival of powerful telescopes including {\em HST}. 
But the comparison of Cepheids and RR\,Lyrae stars remained confined to
LMC, SMC, and one or two additional galaxies 
\citep{Smith:etal:92,vandenBergh:95,Fusi-Pecci:etal:96,Sandage:etal:99}.
\citet{Lee:etal:93}, \citet{Udalski:00}, and \citet{Dolphin:etal:01}
included also the TRGB for comparison \citep[see also][]{Sandage:71}
and the magnitude of the red clump. 
\citet{Sakai:etal:04} extended the comparison of TRGB and Cepheid
distances to 17 galaxies \citep[see also][]{Rizzi:etal:07}.

     The importance of comparing different distance indicators -- i.e.\
mainly Cepheids, RR Lyrae stars and TRGB -- to the level of
$\sim\!0.05\mag$ lies in the new developments which show that the P-L
relation of classical Cepheids differs from galaxy to galaxy. 
It has been argued 
(\citealt*{TSR:03}, hereafter TSR\,03; 
 \citealt*{STR:04}, hereafter STR\,04) 
that the slope and zero point of the Cepheid P-L relation differs
between the Galaxy and LMC, and that the difference is likely to be a
metallicity effect, and may depend on helium as well according to
pulsational models \citep{Marconi:etal:05}.  

The inequality of Cepheids in different galaxies should not come as a
surprise. It was known since \citet{Gascoigne:Kron:65} that SMC
Cepheids are bluer than others, which alone precludes identical P-L
relations. The color difference is not only caused by Fraunhofer
blanketing of the metal lines, but it is also due to a real
temperature difference of Cepheids at given period as was shown already
by \citet{Laney:Stobie:86}. Galaxy-specific differences of Cepheids
were also demonstrated by differences in their light curves at given
period \citep{Tanvir:etal:05,Koen:Siluyele:07}. But it took a wealth
of good data to study galaxy-to-galaxy differences of the P-L relations
themselves, for instance photometry of large numbers of Cepheids in
the Galaxy, LMC, and SMC
\citep{Berdnikov:etal:00,Udalski:etal:99a,Udalski:etal:99b}, 
independently determined reddenings
\citep{Fernie:etal:95,Udalski:etal:99a,Udalski:etal:99b}, and
distances (\citealt{Feast:99}, for Cepheids in Galactic clusters,
\citealt{Fouque:etal:03} and \citealt{Barnes:etal:03} for
moving-atmosphere Baade-Becker-Wesselink [BBW] distances and distances
of LMC and SMC [as compiled from many authors in Tables~\ref{tab:06}
and \ref{tab:07} here]). 
These data show that the P-L relation of Cepheids cannot be
universal. The reason is that the metal-poor LMC Cepheids are shifted
in the luminosity - $T_{e}$ diagram (i.e.\ the instability strip
in the HR diagram) to higher temperatures at constant $L$ as compared
to metal-rich Galactic Cepheids (\citeauthor{STR:04}). 
The different slopes, which the Cepheids define in the instability
strip causes also the {\em slopes\/} of the P-L relations to be
different.   

     The {\em slope\/} difference of the P-L relations of the Galaxy
and LMC (and of short and long-period Cepheids in LMC; see
\citeauthor{STR:04}) is particularly troublesome because the
difference of the absolute magnitude of the Cepheids in the two
galaxies becomes a function of period. While the blue LMC Cepheids are
{\em brighter\/} than their relatively red Galactic counterparts by as
much as $\Delta M_{V}=0.36\mag$ at $\log~P=0.5$, the difference
diminishes with increasing period and changes sign at $\log~P=1.38$. 
From this follows that if the luminosity difference is interpreted as
a metallicity effect, {\em any metallicity correction must depend on
period}. 

     It is a coincidence that the new P-L relations and the
period-dependent metallicity corrections lead to distances, which
{\em on average\/} agree reasonably well with earlier Cepheid
distances, although the latter were based on the unjustified
assumption that the LMC P-L relation of \citet{Madore:Freedman:91} was
universal. As an example, the early Cepheid distances of eight or more
galaxies by \citet{Ferrarese:etal:00}, \citet[][Table~3,
col.~2]{Freedman:etal:01}, and \citet{Tammann:etal:02} agree with
those adopted here and by \citeauthor{STS:06} to within less
than $0.1\mag$, -- regardless whether some kind of bulk metallicity
correction, irrespective of period, is applied and independent of the
adopted LMC zero point (for details see
\citeauthor{STS:06}). 

     However, the near agreement of the old and new Cepheid distances
collapses if Cepheids with non-average properties are considered. For
instance, \citet{Freedman:etal:01} and \citet{Riess:etal:05}, using
the LMC P-L relation of \citet{Udalski:etal:99c} or
\citet{Thim:etal:03}, have based their luminosity calibration of
SNe\,Ia on only six and four galaxies, respectively, whose Cepheids
happen to be particularly metal-rich and to have quite long periods. 
Correspondingly, the new period-dependent metallicity corrections
become important and are the main reason why our present distances
of these galaxies are longer by $\sim\!0.3\mag$ on average than
adopted by these authors (\citealt*{STT:06}; 
\citeauthor{STT:06} hereafter). 
The ensuing discrepancy in $H_{0}$ as derived from Cepheid-calibrated
SNe\,Ia is in the order of 15\%, our scale being longer.

     The purpose of this paper is to construct an independent distance
scale based on Pop.~II stars in order to test our Cepheid
distances. RR\,Lyr magnitudes of 24 galaxies are compiled from the
literature and uniformly reduced (\S~\ref{sec:02:1}). They are
used to calibrate the absolute magnitude $M^{\rm TRGB}_{I}$ of the
TRGB and to test its dependence on metallicity (\S~\ref{sec:02:2}). 
The different P-L relations and their calibration are discussed
in \S~\ref{sec:03}. The (satisfactory) comparison of the Cepheid and
TRGB distances of 18 galaxies is in \S~\ref{sec:04}. 
The Hubble diagrams with increasing outreach from TRGB, Cepheid,
Cepheid-calibrated TF and SN\,Ia distances and the resulting value of
$H_{0}$ are discussed in \S~\ref{sec:05}. The conclusions are in
\S~\ref{sec:06}.

\section{POPULATION II DISTANCE INDICATORS}
\label{sec:02}
%
\subsection{RR Lyrae Stars}
\label{sec:02:1}
%
\subsubsection{The calibration of RR Lyrae stars}
\label{sec:02:1:1}
A summary of many calibration studies of the absolute
magnitude of RR Lyrae stars as function of metallicity has been
given elsewhere \citep*[][hereafter ST\,06]{Sandage:Tammann:06}, the
details of which are not repeated here. However the results are
these.

     (1). It is almost certain that the relation between $M_{V}^{\rm RR}$
and [Fe/H] is non-linear. Most {\em theoretical\/} models of the zero-age
horizontal branch (ZAHB) made after about 1990 predict the
non-linearity. Although the $M_{V}$ luminosity of the level part of
the zero-age HB is a function of [Fe/H] (higher metallicity models
have fainter ZAHB), the variation of $M_{V}$ with [Fe/H] becomes
progressively smaller as [Fe/H] is decreased 
\citep[see Fig.~3 of][reproduced as Fig.~9 of ST\,06]{VandenBerg:etal:00}. 
This non-linearity, of course, applies also to that part of the HB
that contains the RR Lyrae variables. 
Representative theoretical zero-age models are by 
\citet{Lee:etal:90}, 
\citet{Castellani:etal:91},
\citet{Bencivenni:etal:91}, 
\citet{Dorman:92}, 
\citet{Caputo:etal:93},
\citet{Caloi:etal:97}, 
\citet{Salaris:etal:97}, 
\citet{Cassisi:etal:99}, 
\citet{Ferraro:etal:99},
\citet{Demarque:etal:00}, 
\citet{VandenBerg:etal:00}, and 
\citet{Catelan:etal:04}.

     The non-linearity also carries over to the HB that
has evolved away from the ZAHB.

    (2). There are many {\em observational\/} data that also suggest
that the calibration of $M_{V}^{\rm RR}$ with [Fe/H] is non-linear,
many of which are summarized by \citeauthor{Sandage:Tammann:06}. 
Important among these is the analyses of RR\,Lyrae data in many
globular clusters by \citet{Caputo:etal:00}. These authors combine a
pulsation equation that relates period, luminosity, temperature, and
mass with observational data for globular cluster RR\,Lyraes at the
blue edge of the instability strip for overtone pulsators and at the
red edge for fundamental mode pulsators. Their obvious non-linear
$M_{V}$ calibration, shown in their Figure~3, is reproduced as
Figures~11 \& 12 of \citeauthor{Sandage:Tammann:06}. 
Their study using observational data follows earlier non-linear
analyses of $M_{V}^{\rm RR}$ as function of [Fe/H] by 
\citet{Caputo:97}, 
\citet{Gratton:etal:97},
\citet[][their Fig.~15]{DeSantis:Cassisi:99},
\citet{McNamara:etal:04}, and undoubtedly others.
The most recent is the study by \citet{Bono:etal:07} where they show
the non-linearity over the entire range of [Fe/H] from 0 to $-25$ 
(their Fig.~16).

   (3). The zero point of the resulting non-linear $M_{V}^{\rm RR}$
$-$ [Fe/H] relation can be found by several methods, some leading to
the so-called long RR\,Lyrae scale that gives $\langle M_{V}\rangle$
near $+0.52$ at [Fe/H]$ = -1.5$, and the short scale that gives 
$\langle M_{V}\rangle$ near $+0.72$ at [Fe/H]$ = -1.5$.

     There are three calibration methods of high weight that
lead to the long scale. In the order that we assign their
reliability they are these.

     (a) The discovery of Delta Scuti stars in globular
clusters, which are ultra short period dwarf Cepheids whose
population~II prototype is the low metallicity pulsator 
SX\,Phoenicis 
(\citealt{Nemec:89}; \citealt{Nemec:Mateo:90a,Nemec:Mateo:90b};
\citealt{McNamara:97} for reviews),
opened the way for a potentially definitive calibration
of RR\,Lyrae luminosities in globular clusters. There are also a
number of such population~II stars in the nearby field with high-weight
Hipparcos trigonometric parallaxes. Using these as absolute
magnitude calibrators for the globular cluster SX\,Phoenicis stars
gives the distances to globular clusters that contain them, and
hence also the absolute magnitude of the RR Lyrae stars in these
clusters. In this way \citet{McNamara:97} has derived an RR\,Lyrae
calibration that gives 
$\langle M_{V}({\rm RR\,evolved})\rangle = +0.52$ at [Fe/H]$ = -1.5$.

     (b) Main sequence fitting of a globular cluster CM diagram to
the Hipparcos trigonometric parallax data for field subdwarfs of
the appropriate metallicity gives the distance to the cluster. An
extensive literature of the complications of the method exists
(correction for reddening of the main sequences of the globular
clusters, the requirement for precision measurements of [Fe/H]
both for the globular clusters and for the appropriate Hipparcos
subdwarfs, whether the Lutz-Kelker bias correction should be
applied to the Hipparcos parallaxes, etc.), include papers by
\citet{Gratton:etal:97}, \citet{Reid:97,Reid:99}, 
\citet{Carretta:etal:00},
and \citet{VandenBerg:etal:00}. 
Representative studies leading to the long distance scale are by 
\citet{Gratton:etal:97}, \citet{Carretta:etal:00}, 
and \citet{McNamara:etal:04}, among others cited therein. These
calibrations are all consistent with $M_{V}({\rm RR\,evolved})=+0.52$
at [Fe/H]$ = -1.5$ to generally better than $0.05\mag$. A summary by
\citet{Gratton:98} to 1998 is important.

   (c) Several recent models of the ZAHB give $M_{V} = 0.65$
\citep{VandenBerg:etal:00} and $0.60$ \citep{Catelan:etal:04} at
[Fe/H]$ = -1.5$. These must be made $0.09\mag$ brighter to account for
the average evolution away from the ZAHB, which gives a mean $M_{V} =
0.53$ at [Fe/H]$ = -1.5$ for the average RR\,Lyr state. 
Both of these studies give non-linear $M_{V}$(Fe/H) relations.

     $M_{V} = 0.52$ at [Fe/H]$ = -1.5$ is adopted here. This is in
excellent agreement with the RR\,Lyr stars in LMC observed by
\citet{Clementini:etal:03a} at $\langle m^{0}_{V}\rangle=19.06$ after
correction for absorption and an LMC distance modulus of $18.54$
(\citeauthor{TSR:03}, Table~6 with the RR\,Lyr entry removed;
see also Table~\ref{tab:06} below). But it is emphasized that the
RR\,Lyr calibration does not depend primarily on the adopted LMC
distance, which is used in \S~\ref{sec:03:2:2} to calibrate the P-L
relation of the Cepheids in LMC. It is our aim to keep the Pop.~I and
Pop.~II distance scales as independent of each other as possible.

     Combining the parabolic form of the $M^{\rm RR}_{V} - $[Fe/H]
relation of \citet{Sandage:06}, using the pulsation equation together
with the observed $\log P - $[Fe/H] relation for cluster RR\,Lyr
stars, with the adopted value of $M_{V} = 0.52$ at [Fe/H]$ = -1.5$
gives the calibration of 
\begin{equation}
 M_{V}({\rm RR\,evolved}) = 1.109 + 0.600(\mbox{[Fe/H]}) 
                                + 0.140(\mbox{[Fe/H]})^{2}
\label{eq:01}
\end{equation}
over the metallicity range $0>\mbox{[Fe/H]}>-2.5$.
From the way it is derived using the mean evolved level of the
RR\,Lyrae variables in the LMC, equation~(\ref{eq:01}) refers to the
mean evolved absolute magnitude of the variables, not the level of the
zero-age HB at the RR\,Lyrae position on the HB.

     Other methods and analyses lead to a fainter 
RR\,Lyrae calibration. A comprehensive review to 1999 is by
\citet{Popowski:Gould:99}. 
In addition to the methods discussed above, these authors analyze
two methods; (a) globular cluster kinematics where
proper motions are compared with observed radial velocities of
individual cluster stars, and (b), where an observed cluster white
dwarf sequence is fitted to a calibrated HR diagram. Altogether
they discuss seven methods for an RR\,Lyrae calibration keeping the
three that they consider the most robust to be statistical
parallaxes of field RR\,Lyrae, trigonometric parallaxes of field
RR\,Lyrae, and internal cluster kinematics. From these they conclude
that $\langle M_{V}({\rm RR\,evolved})\rangle = +0.71$ at 
[Fe/H]$ = -1.6$. 
This is $0.20\mag$ fainter than equation~(\ref{eq:01}) which gives
$\langle M_{V}({\rm RR\,evolved})\rangle = +0.51$ at [Fe/H]$ = -1.6$.
Among the consequences of this faint calibration is that,
using the observations of the mean level of the RR\,Lyrae in the LMC
similar to but earlier than those of \citet{Clementini:etal:03a},
\citeauthor{Popowski:Gould:99} derive a distance modulus of the LMC as
$\mu^{0}=18.33\pm0.08$. This is $0.21\mag$ smaller than 18.54 which
we take to be correct as shown by Table~6 of
\citeauthor{TSR:03} and Table~\ref{tab:06} here, 
supporting the calibration of equation~(\ref{eq:01}), which we adopt
as our scale in the remainder of this paper.

     The intermediate RR\,Lyrae calibration by \citet{Bono:etal:07}
confirms the non-linear dependence on [Fe/H]. It is based on
convective mixing-length models that give absolute magnitudes averaging
$0.1\mag$ fainter than equation~(\ref{eq:01}) here. Their
equation~(10) gives 
$M_{V}({\rm RR\,evolved}) = +0.64$ at [Fe/H]$ = -1.5$, 
that -- with $\langle V^{0\,{\rm RR}}\rangle =19.06$ for LMC from
sources in Table~\ref{tab:01} -- gives $\mu^{0}(\mbox{LMC})=18.44$,
which is only $0.1\mag$ less than the adopted value in
Table~\ref{tab:06} below.

     We have not discussed the many calibrations of 
$\langle M_{V}^{\rm RR}\rangle$ for individual stars using the 
moving-atmosphere method (BBW) for which
there is a large literature. In the hands of a dozen investigators,
these RR\,Lyr calibrations cover the range of the short and long
scale values from $\langle M_{V}^{\rm RR}\rangle$ of $+0.7$ to
$+0.5\mag$ at [Fe/H]$=-1.5$, and therefore are of no help here to
decide between them at the $0.2\mag$ level.

\subsubsection{Twenty-four RR Lyr distances to nearby galaxies}
\label{sec:02:1:2}
The purpose of this section is to summarize the recent data
on detection and measurement of the RR\,Lyr variables in nearby
galaxies and to use equation~(\ref{eq:01}) to determine RR\,Lyr star
distances to them.

    The literature has been surveyed up to the end of 2006. The
results are given in Table~\ref{tab:01} ordered by right ascension. 
Column~(2) shows the number of RR\,Lyr in the particular study. The
metallicity given by the original authors is in column~(3). 
Columns~(4) and (5) list the measured $\langle V\rangle$ and the
$E(B\!-\!V)$ reddenings from \citet{Schlegel:etal:98}. A few of the
papers used the $g$  photometric band \citep{Thuan:Gunn:76} rather
than $V$. These were transformed to $V$ by $\langle V\rangle - \langle
g\rangle = +0.04$  \citep{Saha:etal:90}. 
Column~(6) is the assumed $A_{V}$ absorption 
calculated from $3.1 \times E(B\!-\!V)$. The absorption-free $V^{0}$
values, found by combining columns~(4) and (6) are in column~(7). The
calculated absolute $\langle M_{V}({\rm RR\,evolved})\rangle$
magnitude of the RR\,Lyr stars using equation~(\ref{eq:01}) is in
column~(8).  
The value used by the original authors based on their various
adopted RR\,Lyr calibrations is in column~(9). Column~(8)
combined with column~(7) gives the adopted distance modulus in
column~(10). The distance from the original authors in column~(11) is
not necessarily the difference of columns~(7) and (9) mainly because
of differences of the adopted absorption. The telescope used for the
literature study is in column~(12). The literature reference is in
column~(13), identified as a footnote to the Table, with the details
in the References. 

     Our adopted $\mu^{0}_{\rm new}$ values in column~(10) are the
basis for the distance scale to which all other scales are compared in
the remainder of this paper.
These adopted RR~Lyrae distances agree to $0.04\pm0.08$ with those
published by the original authors, or $0.02\pm0.01$ if the early
determinations for NGC\,147 and NGC\,185 are neglected. We take this
good agreement as a broad consensus with our equation~(\ref{eq:01})
and our adopted RR~Lyrae distance scale.


\subsection{The Tip of the Red-Giant Branch (TRGB)}
\label{sec:02:2}
The potential of the infrared TRGB magnitude as a distance indicator
has been pioneered by \citet{DaCosta:Armandroff:90}. The basis of
their work is that the cores of red giants with initial masses
$\la2\,{\mathfrak M}_{\sun}$ are fully degenerate at the moment when
the helium flash occurs and with nearly constant core masses their
luminosity increases only mildly with increasing $Z$
\citep{Rood:72,Sweigart:Gross:78}. This increase of $M_{\rm bol}$ is
compensated in the $I$-band by the increasing effect of line
blanketing such that $M^{\rm TRGB}_{I}$ becomes a useful standard
candle for old, {\em metal-poor\/} populations.
The importance of this is that $M^{\rm TRGB}_{I}$ extends the Pop.~II
distance scale by a factor of $\sim\!10$ beyond the reach of RR\,Lyr
stars. Moreover a great wealth of apparent $m^{\rm TRGB}_{I}$
magnitudes has since been accumulated by many authors.

     The method of the TRGB is generally employed only in the range
$-2.3< \mbox{[Fe/H]}<-0.7$ because more metal-rich red giants have an
increasingly fainter tip magnitude (see below). If it is restricted to
populations older than $7\;$Gyr the effect of age is negligible
\citep{Lee:etal:93,Rejkuba:etal:05}.

\subsubsection{The calibration of $M^{\rm TRGB}_{I}$}
\label{sec:02:2:1}
The 24 galaxies with RR\,Lyr distances in Table~\ref{tab:01} are
repeated in Table~\ref{tab:02}, where also the corresponding apparent
$m^{\rm TRGB}_{I}$ magnitudes (column~4) and their references
(column~6) are given. The resulting absolute values\footnote{The
  values of $M_{I}^{\rm TRGB}$ and $m_{I}^{\rm TRGB}$ are corrected
  for Galactic absorption throughout.} of 
$M^{\rm TRGB}_{I} = m^{\rm TRGB}_{I} - \mu^{0}_{\rm RR}$ are given in
column~(5). The mean of the absolute magnitudes $M^{\rm TRGB}_{I}$,
omitting Sag dSph whose TRGB is not well defined, and Phoenix whose
RR\,Lyr distance is uncertain, is
\begin{equation}
 M^{\rm TRGB}_{I} = -4.05 \pm 0.02,
\label{eq:02}
\end{equation}
which we adopt. The value holds for a mean TRGB color of
$(V\!-\!I)^{\rm TRGB}=1.6$ (see Table~\ref{tab:02}, col.~2), which
translates into [Fe/H]$\,\sim\!-1.5$ (see below).
The standard deviation of the individual determinations of
$M^{\rm TRGB}_{I}$ is $\sigma=0.08\mag$. This is smaller than expected
from observational errors alone. It follows from this that the random
error of a single RR\,Lyr star {\em or\/} TRGB distance is in any
case smaller than $0.1\mag$ even if metallicity corrections of the
TRGB are neglected (see \S~\ref{sec:02:2:3}).   


     The six late-type galaxies in Figure~\ref{fig:01} deviate
from the zero-line by $0.04\pm0.03\mag$, being brighter. 
The near agreement of $M^{\rm TRGB}_{I}$ for late-type galaxies
and for dwarf spheroidals indicates that internal absorption in the
parent galaxy is negligible for all practical purposes.

\subsubsection{Other calibrations of $M^{\rm TRGB}_{I}$}
\label{sec:02:2:2}
\citet{DaCosta:Armandroff:90} have based their TRGB calibration on
globular clusters with RR\,Lyr distances; they have obtained
$M^{\rm TRGB}_{I}=-3.98$. From the same method \citet{Sakai:etal:04}
have adopted $-4.05$. 
The value of $-4.06$ of \citet{Ferrarese:etal:00b} was calibrated by
Cepheids. 
The luminosity of $-4.07$ by \citet{Bellazzini:etal:04a} rests on the
distances of $\omega$ Cen and 47 Tuc.
\citet{Salaris:Cassisi:97} have determined the TRGB magnitude from
theoretical stellar evolution models and found $-4.16$, which however
includes also very short-lived stars.
They have revised \citeyearpar{Salaris:Cassisi:98} their result to
$-4.27$ and $-4.24$, 
respectively, depending on the adopted bolometric correction. The
result was closely confirmed by \citet{Rejkuba:etal:05} based on
the stellar evolution database of \citet{Pietrinferni:etal:04}.
\citet[][Fig.~15]{Bergbusch:VandenBerg:01} imply a value close to
$-4.05$ based on the models of \citet{VandenBerg:etal:00}.
\citet{Rizzi:etal:07} have fitted the HB of five galaxies to
the metal-dependent HB of \citet{Carretta:etal:00} whose zero point
rests on trigonometric parallaxes; their result is $-4.05\pm0.02$. All
values of $M^{\rm TRGB}_{I}$ quoted in the Section refer to a
metallicity of [Fe/H]$=-1.5$. The calibration with RR\,Lyr stars in
\S~\ref{sec:02:2:1}, which refers to the same metallicity, is in good
to excellent agreement with the values quoted here.

\subsubsection{The dependence of $M^{\rm TRGB}_{I}$ on metallicity}
\label{sec:02:2:3}
\begin{figure}[t]
   \epsscale{0.6}
   \plotone{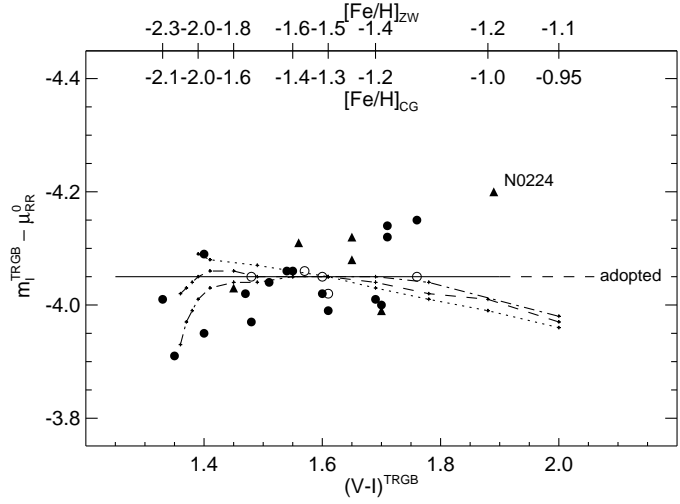}
   \caption{Absolute TRGB magnitudes $M_{I}^{\rm TRGB}$, as
   determined from the difference of the apparent TRGB magnitude and
   the RR\,Lyr star distance, in function of the color $(V\!-\!I)$ and
   metallicities of the TRGB. The corresponding metallicities are
   given at the upper edge of the figure (see text). 
   Note that bluewards of $(V\!-\!I)=1.4$ the color becomes
   insensitive to metallicity. The six late-type galaxies are shown as
   triangles. The five independent calibrators of
   \citet{Rizzi:etal:07} are shown as open symbols. 
   Semi-theoretical predictions of the dependence of $M^{\rm
   TRGB}_{I}$ on metallicity of three different groups are
   drawn; they are normalized to $M^{\rm TRGB}_{I}=-4.05$ at
   $(V\!-\!I)=1.6$ 
   (\citealt{Salaris:Cassisi:98}, eq.~(5): dashed; 
   \citealt{Bellazzini:etal:04a}: dashed-dotted;
   \citealt{Rizzi:etal:07}: dotted).}
\label{fig:01}
\end{figure}

The absolute magnitudes $M^{\rm TRGB}_{I}$ of Table~\ref{tab:02} are
plotted against the color $(V\!-\!I)$ of the TRGB in
Figure~\ref{fig:01}. The colors are taken from the original
literature. They are converted into metallicities [Fe/H]$_{\rm ZW}$ on
the scale of \citet{Zinn:West:84} following \citet{Bellazzini:etal:01}
and shown at the upper edge of Figure~\ref{fig:01}. The metallicities
[Fe/H]$_{\rm CG}$ in the system of \citet{Carretta:Gratton:97} are
also shown.

     The calibrators in Figure~\ref{fig:01} suggest an increase of the
TRGB luminosity with increasing metallicity, the reality of which we
doubt however. The five calibrators of \citet{Rizzi:etal:07} give a
flat calibration, although for a narrower metallicity rang of
$-1.8<\mbox{[Fe/H]}_{\rm ZW}<-1.3$. Also the models of
\citet{VandenBerg:etal:00} do not show a systematic change of the
TRGB with metallicity
\citep[][Fig.~13]{Bergbusch:VandenBerg:01,Rejkuba:etal:05}; the tip
becomes {\em fainter\/} only for the most metal-rich red giants with
$(V\!-\!I)>2.0$. The strong decline of the tip magnitude redwards of
this limit has been directly observed in rich populations with a wide
metallicity spread as in the Galactic bulge \citep{Zoccali:etal:03}
and in the halo of NGC\,5128 \citep{Rejkuba:etal:05}.

     Model-dependent variations of the TRGB magnitude with metallicity
have also been determined by \citet{Salaris:Cassisi:98},
\citet{Bellazzini:etal:04a}, and \citet{Rizzi:etal:07}. Their results
are displayed in Figure~\ref{fig:01}, after they are normalized to
$M^{\rm TRGB}_{I}=-4.05$ at $(V\!-\!I)=1.6$. The authors agree that
the TRGB magnitude does not change by more than $0.05\mag$ over the
interval $1.4<(V\!-\!I)\la1.9$ or $-2.0<\mbox{[Fe/H]}_{\rm
  ZW}<-1.2$. For the most metal-poor red giants the results
diverge. For the metal-rich red giants with [Fe/H]$_{\rm ZW}>-1.2$ the
results agree on a progressive dimming of $M^{\rm TRGB}_{I}$.

     The near constancy of $M^{\rm TRGB}_{I}$ over a wide metallicity
interval is fortunate because no metallicities or colors
$(V\!-\!I)^{\rm TRGB}$ are available for most galaxies with known TRGB
magnitudes.

     In the following a constant value of $M^{\rm TRGB}_{I}=-4.05$
will be adopted irrespective of metallicity, as
\citet{Ferrarese:etal:00b} as well as
\citet{Karachentsev:etal:04,Karachentsev:etal:06,Karachentsev:etal:07}
in their extensive work on the TRGB have done. Since metal-rich giant
branches are unfrequent in old populations the mean distance error
incurred will hardly be larger than $0.05\mag$.
If one compares the distances of 22 galaxies,
for which \citet{Rizzi:etal:07} give metal-corrected TRGB distances
with those one obtains from a fixed calibration of 
$M_{I}^{\rm TRGB}=-4.05$ they differ by only $0.03\mag$ with a
small standard deviation of $0.06\mag$. A similar conclusion is
reached below where TRGB distances with and without metallicity
corrections are compared with (metallicity-corrected) Cepheid
distances (Table~\ref{tab:09}).

\subsubsection{TRGB distances of field galaxies}
\label{sec:02:2:4}
\citet{Karachentsev:etal:04} have compiled many $m_{I}^{\rm TRGB}$
magnitudes and have provided additional ones in
\citet{Karachentsev:etal:06,Karachentsev:etal:07}. 
Other authors have observed the TRGB in many additional galaxies. 
Altogether, $m_{I}^{\rm TRGB}$ magnitudes are available for
218 (mostly dwarf) galaxies. Since \citeauthor{Karachentsev:etal:04}
have used the same TRGB calibration as adopted here, their listed
distances remain unchanged.
For consistency all distances have been slightly adjusted,
where necessary, to the present calibration of $M_{I}^{\rm TRGB}=-4.05$.

\subsubsection{ TRGB, the Virgo cluster, and the cosmic distance scale}
\label{sec:02:2:5}
\citet{Caldwell:06} has observed the TRGB in the Virgo galaxy NGC\,4407
and in five small dwarf galaxies in its vicinity and obtains -- with
our calibration -- a mean distance of $\mu^{0}=31.08\pm0.05$. An
anonymous Virgo dwarf away from other galaxies yields $31.22$ with a
more realistic error of $\pm0.17$ \citep{Durrell:etal:07}. In view of
the depth effect of the Virgo cluster, which amounts to 2-3$\;$Mpc 
\citep[eg.][]{Mei:etal:07} even on the assumption of sphericity, 
these first distances to individual cluster members cannot be taken as
giving the mean cluster distance. TRGB distances of a statistically
fair sample of Virgo cluster members will be most valuable as a test
for the entire distance scale.   

     TRGB stars have also been detected in the intracluster medium of
the Virgo cluster. \citet{Durrell:etal:02} and \citet{Caldwell:06}
quote distances of $31.36^{+0.27}_{-0.17}$ and $31.2\pm0.09$,
respectively.\footnote{\citeauthor{Durrell:etal:02} do not actually
  quote a distance to their Virgo cluster fields. The value
  $\mu^{0}=31.36$ follows from their $I^{\rm TRGB}=27.31$ and the
  calibration of $M_{I}=-4.05$ used here.} 
However, these are only {\em lower\/} limits to the
distance of the cluster core, because only the nearest TRGB stars can
be detected in the cluster field, while the more distant ones are
drowned among the red-giant stars on the {\em near\/} side. 

     Information on the TRGB distances is available for four SNe\,Ia at
present. Their relevant data are set out in Table~\ref{tab:03}.
The corrected $m_{V}(\max)$ magnitudes of the SNe\,Ia in column~(2)
are from \citet*[][Table~2, column~9, in the following
RTS\,05]{RTS:05}. 
The TRGB distances of the individual parent galaxies (column~4) and of
the mean TRGB distances of their respective groups (column~6),
together with the number of group members involved (column~7), are
from the sources identified in column~(8). 
All distances are based on $M_{I}^{\rm TRGB}=-4.05$. 
The resulting absolute magnitude of the four SNe\,Ia, based on the
group distances, is given in column~(9). The mean absolute magnitude
of $M_{V}(\mbox{SNe\,Ia})=-19.37\pm0.06$ is statistical the same as
$-19.46\pm0.04$, which is based on 10 SNe\,Ia with Cepheid distances
(\citeauthor{STS:06}) and which has much higher weight. 
The latter value is also close to various values derived by others, as
summarized by \citet{Gibson:etal:00} in the $B$-band. (Note that
$B^{0}_{\max}-V^{0}_{\max}=-0.03$; \citeauthor{STS:06}).
When the TRGB method can be pushed to yield reliable 
distances out to $\sim\!31.5$, seven presently known SNe\,Ia will come
into its reach and will yield an {\em independent\/} calibration of
$H_{0}$ through the TRGB-calibrated Hubble diagram of SNe\,Ia.


\section{POPULATION~I DISTANCES}
\label{sec:03}
The foundation of the Population~I distance scale is classical
Cepheids. Their metallicity-dependent period-luminosity (P-L)
relations in $B$, $V$, and $I$ have been derived in
\citeauthor{TSR:03} and \citeauthor{STR:04}. It was  
found that the relatively metal-rich Cepheids in the Solar neighborhood
([O/H]$_{\rm Te}=8.50$ in the $T_{e}$-based scale of 
\citet{Kennicutt:etal:03} and \citet{Sakai:etal:04}) define a P-L
relation that differs in slope and shape from the P-L relation of LMC
([O/H]$_{\rm Te}=8.34$). It is therefore not possible to determine a
LMC distance from a P-L relation that is based on {\em Galactic\/}
Cepheids. The P-L relations of the two galaxies must be independently
be zero-pointed. A more general discussion of the problem follows.

\subsection{The forms of the P-L relation}
\label{sec:03:1}
The only rational to assume that the P-L relation is universal is
convenience since the time in which it was known that Cepheids in
different galaxies have different colors, temperatures, light curves,
and slopes of their P-L relations (see \S~\ref{sec:01}).
Also the break of the P-C and P-L relations of LMC at $P=10^{\rm d}$,
not yet seen in other galaxies, is alarming (see below).
However, the investigation of the shape of the P-L relation of
individual galaxies is difficult because the intrinsic width of the
instability strip requires {\em very\/} large Cepheid samples
distributed over a wide period interval. 
Such samples are available only for LMC and SMC; they will never
become attainable in dwarf galaxies. One has therefore to assume, in
first approximation, that the P-L relations are {\em linear}.

     Even on the assumption of linearity the determination of the 
{\em slope\/} is demanding for several reasons.
(a) The Cepheids in many galaxies, particularly the distant ones, are
often restricted to $P\ga10^{\rm d}$.
(b) Selection bias in favor of Cepheids with short periods near to the
detection limit \citep{Sandage:88} tends to flatten the slope.
(c) The slope is independent of the reddening only as long as it does
not depend on the period, which is not warranted a priori.

     The reddening values $E(B\!-\!V)$ of the Cepheids in all galaxies
considered are derived from $(V\!-\!I)$ and in some cases $(B\!-\!V)$
colors and an {\em adopted\/} template P-C relation with the exception
of only three galaxies, viz. the Galaxy, LMC, and SMC.
The individual reddenings of the Galactic Cepheids have been derived
{\em ab initio\/} by \citet{Fernie:90}, \citet{Fernie:etal:95}, and
other authors. They have been homogenized and slightly revised by
\citeauthor{TSR:03}. The reddenings of the Cepheids in LMC
and SMC have been determined from adjacent red-clump stars by
\citet{Udalski:etal:99a,Udalski:etal:99b}.

\subsubsection{The shape of the Galactic P-L relation}
\label{sec:03:1:1}
The shape of the P-L relation of the metal-rich Galactic
Cepheids ([O/H]$_{\rm Te}=8.50$) is determined from two independent
methods both covering a wide period interval. 

(a) Thirty-three Cepheids in clusters are taken from the revised list
of \citep[][see STR\,04]{Feast:99}. Their distances are known from
main-sequence fitting relative to the Pleiades whose distance modulus
of $\mu^{0}=5.61\pm0.02$ is well determined from different
methods, including trigonometric parallaxes 
(\citeauthor{STR:04}).

(b) BBW distances are
available for 33 partially overlapping Galactic Cepheids 
from \citet{Fouque:etal:03} and \citet{Barnes:etal:03}. Also 
included are three additional Cepheids with distances from
interferometric diameter measurements 
\citep{Benedict:etal:02,Nordgren:etal:02,Lane:etal:02,Kervella:etal:04}.

     The Cepheids under (a) and (b) are corrected for Galactic
absorption. The absorption-free magnitudes of
the two sets of Cepheids define P-L relations in $B$, $V$ and $I$ with
very similar slopes. While they agree at $\log P = 0.5$ to within
$0.01\mag$, they diverge in all three colors by not more than 
$0.2\mag$ at $\log P=1.5$. The adopted mean linear P-L relations in
\citeauthor{TSR:03}, equations~(16)--(18), 
should therefore be good to within $\pm0.1\mag$ over a wide period
interval. 

\begin{figure}[t]
   \plotone{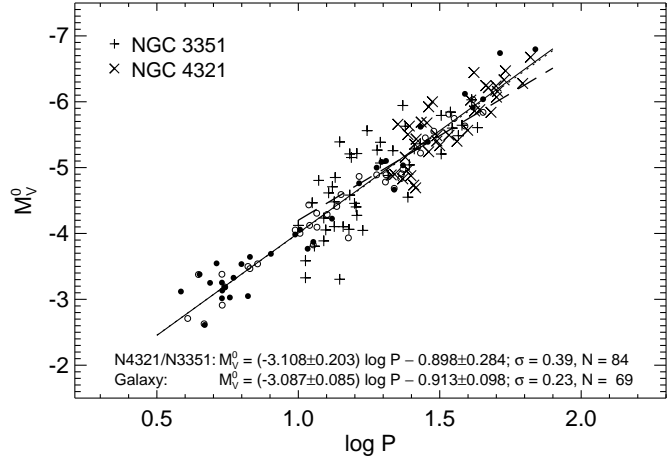}
   \caption{P-L relation in $V$ of metal-rich Cepheids in the
   Galaxy ({\em circles}), NGC\,4321 ($\times$), and NGC\,3351 
   ({\em crosses}). The latter two galaxies define a slope in good
   agreement with the Galaxy ({\em dotted line}). For comparison the 
   LMC P-L relation for $\log P>1.0$ is shown as a dashed line.} 
\label{fig:02}
\end{figure}

     The Galactic P-L relation with slope $-3.087\pm0.085$ is as
linear in all three colors as can be determined from a sample of only
69 Cepheids. Provided that the values of $E(B\!-\!V)$ 
from \citeauthor{Fernie:etal:95} do not systematically overestimate the
reddening of long-period Cepheids -- a possibility which has been
discarded in \citeauthor{TSR:03} -- the Galactic P-L relation
does not have the break at $\log P=1.0$ as is observed in LMC (see
below). The Galactic P-L relation in $B$, $V$, and $I$ is steeper than
observed in most other galaxies, but the slope is about equally steep
in the metal-rich galaxies NGC\,3351 and NGC\,4321 (Fig.~\ref{fig:02})
as well as in NGC\,224 (M\,31; \S~\ref{sec:03:3} and \S~\ref{sec:03:1:5}).


     From trigonometric parallaxes with the fine guidance sensor on
{\em HST\/} \citet{Benedict:etal:07} have derived a very flat slope of
the Galactic P-L relation of $-2.46$ in $V$, but of their 10 Cepheids
only l~Car has a period significantly larger than 10 days. The flat
relation raises several questions. (1) The authors discuss the
possibility of a break of the Galactic P-L relation. But this creates
more problems than it solves. (2) The flat P-L relation predicts 18
Cepheids in Galactic clusters with periods $0.6<\log P<1.0$ to be
brighter by $0.16\pm0.05$ than listed in Table~1 of
\citeauthor{TSR:03}. These Cepheids lie all in well defined
clusters (not in less reliable associations!), and an {\em upward\/}
revision of the cluster distance scale by this amount is difficult to
accept, particularly since \citet{An:etal:07} have concluded from
refined photometry of 7 of the 18 clusters that their distances, if
anything, should be shifted {\em downwards\/} by $0.12\pm0.06$. (3)
The luminosity difference between the 10 Cepheids by
\citet{Benedict:etal:07} and the P-L relation adopted here is not
only a function of period, but also a function of {\em apparent\/}
magnitude. This opens the possibility of astrometric errors in
function of magnitude. Finally the 10 parallax Cepheids define a P-L
relation in $V$ with a random scatter of only $0.11\mag$ as compared
to $0.22\mag$ in LMC. This confirms the prediction that the intrinsic
half-width of the Galactic P-L relation is only 0.08 on the basis of
the flat constant-period lines in the Galaxy
(\citeauthor{STR:04}). Alone the slope difference of the
constant-period lines between the Galaxy and LMC constitutes an
important difference between the Cepheids of these two galaxies.

     The {\em HST\/} parallaxes by \citet{Benedict:etal:07} have been
augmented by Hipparcos parallaxes including four additional
Cepheids by \citet{vanLeeuwen:etal:07}. These authors have derived a
P-L relation from a combination of $V$ and $I$ magnitudes. Since this
may conceal differences of the separate P-L relations, the relation in
$V$ was derived from their data after correcting for absorption. Not
surprisingly the resulting relation is essentially the same as by
\citet{Benedict:etal:07} because most of the weight lies on the 
{\em HST\/} parallaxes.

     \citet{Fouque:etal:07} rely for the slope of the Galactic P-L
relation on 49 BBW infrared surface brightness distances, augmented by
the 10 trigonometric parallaxes of \citet{Benedict:etal:07}; they
exclude Cepheid distances from open clusters. In this way they derive
a slope in $V$ of $-2.678$, significantly flatter than our value of
$-3.087$ and close to the slope one obtains if the LMC Cepheids are
(unjustifiedly) fitted with a {\em single\/} slope. The authors admit
that the crux of the BBW method is the correct choice of the $p$-factor
that converts observed radial velocities into pulsational
velocities. They have taken $p$ from the model of
\citet{Nardetto:etal:04}, where $p$ depends on the period $P$. This is
unfortunate because any error of the dependence of $p(P)$ translates
into an error of the slope of the P-L relation. A weaker $p(P)$
dependence yields steeper P-L relations. 

     The slope difference between \citet{Fouque:etal:07} and us
(\citeauthor{STR:04}) is caused by their almost exclusive
reliance on the BBW method, while we rely on cluster distances which
agreed impressively well with the BBW distances available at the time
from \citet{Fouque:etal:03} and \citet{Barnes:etal:03}. Our steep
slope finds support in metal-rich Cepheids of other galaxies with
define, as discussed in \S~\ref{sec:03:1:5}, an equally steep slope. 

     If \citet{Benedict:etal:07} and \citet{Fouque:etal:07} claim that
the Galactic and LMC P-L relations are undistinguishable, they
fail to acknowledge the {\em break\/} of the LMC relation at $\log
P=1.0$ (see \S~\ref{sec:03:1:2}) and its absence in the Galactic
Cepheids. In addition the inequality of the Galactic and LMC P-L
relations will be shown in \ref{sec:03:1:4}, independent of any
adopted distances, only on the basis of the Cepheid colors. 

     The pulsation models of \citet{Marconi:etal:05} for
high-metallicity Cepheids ($Z=0.02$) do not give as steep a slope as
we observe in the Galaxy. They obtain the steepest slope for $Y=0.26$
with flatter slopes for higher $Y$ (0.28, 0.31) {\em and\/} lower $Y$
(0.25), but even in the first case the slope is significantly flatter
than observed. Surprisingly, lower-metallicity models with $Z=0.01$,
$Y=0.26$, a composition actually favored for $\delta\;$Cep by
\citet{Natale:etal:07}, come close to the observed slope for the Galaxy. 
Yet the model slopes are not yet definitive because they depend on the
position of the red edge of the instability strip, where the treatment
of convection is necessary. Also the uneven population of the strip
due to temperature-dependent crossing times should be accounted for.
Furthermore, the pulsation models show that the P-L relation depends
not only on $Z$ but on $Y$ as well.  
The point is that the models of \citet{Marconi:etal:05} do show that
the P-L slopes should vary from galaxy to galaxy.

\subsubsection{The shape of the LMC P-L relation}
\label{sec:03:1:2}
\begin{figure}[t]
   \epsscale{0.9}
   \plotone{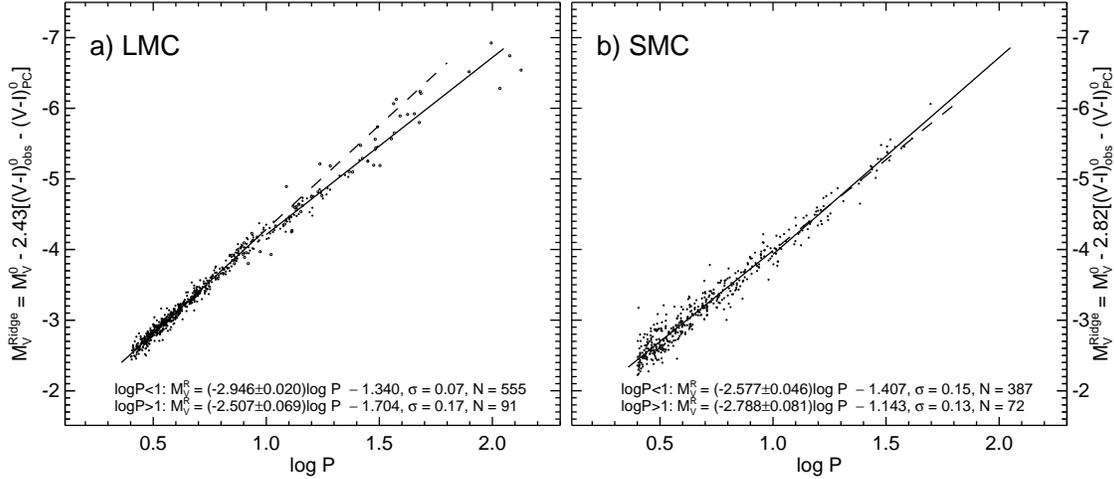}
   \caption{Left panel: Ridge line P-L relation in $V$ for
     LMC. The break at $\log P=1.0$ is highly significant. 
     Right panel: Ridge line P-L relation in $V$ for SMC
     omitting Cepheids with $\log P<0.4$. The dashed lines are the
     extrapolations of the P-L relation of the Cepheids with 
     $\log P<1.0$.}  
\label{fig:03}
\end{figure}

The shape of the P-L relation in $B$, $V$, and $I$ of the
low-metallicity Cepheids of LMC ([O/H]$_{\rm Te}=8.36$) is unusually
well determined by about 680 Cepheids from the OGLE program
\citep{Udalski:etal:99a} and several other sources
(\citeauthor{STR:04}). A linear fit over the entire period
interval with slope $-2.702$ in $V$ is not the optimum fit. A
significantly better fit is achieved by two linear lines breaking at
$P=10^{\rm d}$ \citep{Tammann:Reindl:02,Tammann:etal:02,Ngeow:etal:05}. 
The break, also clearly seen in the P-C relations for $(B\!-\!V)$ and
$(V\!-\!I)$ (\citeauthor{STR:04}, Fig.~1a, 1b), withstands
several statistical tests
\citep{Ngeow:etal:05,Kanbur:etal:07,Koen:etal:07}.
The break becomes particularly striking if the Cepheids are reduced to
the P-L ridge line by shifting them along constant period lines.
The shift is determined by the difference between the observed color
$(V\!-\!I)^{0}_{\rm obs}$ of a Cepheid with fixed period and the color
$(V\!-\!I)^{0}_{\rm PC}$ required by the appropriate P-C relation for
this period, i.e.\ 
\begin{equation}
 M_{V}({\rm Ridge}) = M^{0}_{V} - \beta_{V,V\!-\!I}[(V\!-\!I)^{0}_{\rm
 obs} - (V\!-\!I)^{0}_{\rm PC}]. 
\label{eq:03}
\end{equation}
The coefficient $\beta_{V,V\!-\!I}$ is the slope of the
constant-period lines in the CMD for $M_{V}$ versus $(V\!-\!I)$. For
LMC it was found $\beta_{V,V\!-\!I}=2.43$
(\citeauthor{STR:04}, eq.~[29]).
The resulting P-L relation with its clear break is shown in
Figure~\ref{fig:03}.


     The pulsational models for $Z=0.004$, $Y=0.25$ of
\citet{Marconi:etal:05} fit the observed P-L relation of LMC 
well, {\em including\/} the break at $10^{\rm d}$. The theoretical
break is even more pronounced than observed.

\subsubsection{The shape of the SMC P-L relation}
\label{sec:03:1:3}
Linear regressions to 459 SMC Cepheids of
\citet{Udalski:etal:99b} in the range $0.4 < \log P < 1.7$ give 
\begin{eqnarray}
 \label{eq:05}
   M^{0}_{B} & = & -(2.222\pm0.054)\log P - (1.182\pm0.041), \\
 \label{eq:06}
   M^{0}_{V} & = & -(2.588\pm0.045)\log P - (1.400\pm0.035),
   \quad \mbox{and} \\
 \label{eq:07}
   M^{0}_{I} & = & -(2.862\pm0.035)\log P - (1.847\pm0.027).
\end{eqnarray}
The constant terms in equations~(\ref{eq:05})--(\ref{eq:07})
are based on the distance of $\mu^{0}_{\rm SMC}=18.93$ from 
Table~\ref{tab:07} below. The determination of the exact shape of the
SMC P-L relations, however, is subtle. It is known that they turn
downwards for the many SMC Cepheids with very short periods ($\log
P<0.4$; \citealt{EROS:99}). It is difficult to decide whether this
should be interpreted as a break at $\log P \sim 0.4$ or as curvature
of the P-L relations. The P-C relations in $(B\!-\!V)$ and $(V\!-\!I)$
clearly suggest an additional break at $\log P =1.0$ like in LMC. 
The ridge line P-L relation in $V$, constructed with
$\beta_{V,V\!-\!I}=2.82$ appropriate for SMC
(\citeauthor{STR:04}, Table~4), also suggests the break at
$\log P=1.0$ (at only a $2\sigma$ level), but -- contrary to LMC --
with the slope increasing above the break point (Fig.~\ref{fig:03}b).
The single-slope SMC P-L relations of
equations~(\ref{eq:05})--(\ref{eq:07}), however, were deemed to be
adequate for the following application to very metal-poor Cepheids
(see \S~\ref{sec:03:3} below).

\subsubsection{The interplay of the P-L and P-C relations}
\label{sec:03:1:4}
The ongoing discussion on the slope of the Galactic P-L relation could
still nourish the hope that the P-L relations of classical Cepheids
were universal. This hope is unfounded in view of the 
period-{\em color} (P-C) relations. Because if the $B,V,I$ P-L
relations are to be invariable, so must be the P-C relations in (B-V)
and (V-I), which are simply the differences of the corresponding P-L
relations. 

     Yet the metal-poor LMC and even more metal-poor SMC Cepheids are
on average {\em significantly\/} bluer in $(B\!-\!V)^{0}$ than the
Galactic Cepheids by 0.07 and $0.14\mag$, respectively. The
corresponding number for $(V\!-\!I)^{0}$ is $0.05\mag$ for both
galaxies. Thus the zero points of the $B,V,I$ P-L relations must
differ by at least this amount, but the shift could be larger by any
additional constant amount. The color behavior of the Cepheids in a
two-color diagram $(B\!-\!V)^{0}$ versus $(V\!-\!I)^{0}$ can be
explained -- neglecting their periods --  by atmospheric models
\citep{Sandage:etal:99} as the blanketing effect of the metal lines 
(see \citeauthor{TSR:03}, Fig.~7a).

     {\em But\/} in addition the same models show LMC Cepheids to be
hotter than Galactic Cepheids {\em at given period\/} by roughly 
200$\;$K \citep[see also][]{Laney:Stobie:86} and to be also hotter at
constant luminosity (\citeauthor{STR:04}, Fig.~20). This is, as has
been shown, an even stronger luminosity effect than the line
blanketing.

     If the size of the blanketing effect and of the temperature
difference were independent of period, the {\em slopes\/} of the P-L
relations could still be the same everywhere, and only their
zero points were shifted. However, it is clear that the blanketing
effect depends on color and hence on period. Moreover it was shown in
\citeauthor{STR:04} that also the temperature difference at
constant luminosity increases with period. These effects {\em must\/}
reflect on the slopes of the P-C relations. In Table~\ref{tab:04} the
observed slopes of the P-C relations in $(B\!-\!V)^{0}$ and 
$(V\!-\!I)^{0}$ of the Galaxy, LMC, and SMC are compiled. To avoid
further complications with the break at 10d of at least the LMC
relations, the slopes in the interval $0.4\le\log P\le 1.0$ are only
considered. The slope differences in Table~\ref{tab:04} between the
three galaxies are highly significant at the 2-5$\sigma$ level.


     The conclusion is that since the P-C relations have different
slopes in galaxies with different metallicity, the slopes of the
$B,V,I$ P-L relations must also vary. It is therefore not anymore the
question whether the P-L relations are universal, but only by how much
they vary.

     A word of warning may here be in place. There are indications
that the P-L relations in the near infrared ($JHK$) have closely the
same slope independent of metallicity. If this is the case, this still
does not mean, as discussed above, that they have the same
zero point. It will therefore be necessary to independently zero point
the near infrared P-L relations for Cepheids with different chemical
composition.

\subsubsection{The slope of the P-L relation in function of metallicity}
\label{sec:03:1:5}
The metallicities (from \citealt{Sakai:etal:04}) and slopes of the
$B$, $V$, and $I$ P-L relations of nine galaxies are compiled in
Table~\ref{tab:05}. Only galaxies with well or reasonably well
determined slopes are considered. The original sources of the Cepheid
data are listed in the last column.


     The Cepheids in Sextans A and B are combined to a single P-L relation   
because they have nearly the same (very low) metallicity and almost
identical TRGB distances ($\mu^{0}=25.78,25.79$; \citealt{Rizzi:etal:07}). 

\begin{figure}[t]
   \epsscale{0.6}
   \plotone{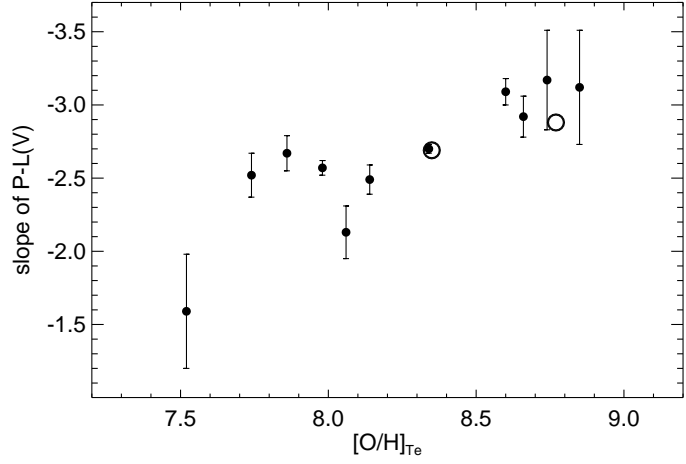}
   \caption{Slope of various P-L relations in $V$ in function of the
   metallicity [O/H]$_{\rm Te}$. Open circles are the
   means of seven galaxies each from \citeauthor{STT:06} (Fig.~10).} 
\label{fig:04}
\end{figure}

     The decrease of the P-L slope with decreasing metallicity in
Figure~\ref{fig:04} is striking. The extreme case of Sextans A and B
deserves special emphasis. 
Confirmatory work would be interesting, although the
two small galaxies may not have many more Cepheids than already known
(17 over a wide period interval). It is likely that some of the
scatter in Figure~\ref{fig:04} is intrinsic. The available data for
several metal-rich galaxies admittedly suggest that their P-L relations
are flatter than in the Galaxy (see \S~\ref{sec:03:4:4}). But
Figure~\ref{fig:04} leaves no doubt that the P-L slope {\em is\/}
correlated with [O/H]. Hence, the P-L relation cannot be universal but
must vary from galaxy-to-galaxy, primarily as a function of [O/H].


\subsection{The zero-point calibration of the P-L relation of Cepheids}
\label{sec:03:2}
%
\subsubsection{The zero point of the Galactic P-L relation}
\label{sec:03:2:1}
The zero point of the Galactic P-L relation for an adopted metallicity
of [O/H]$_{\rm Te}=8.62$ rests on 33 cluster distances
(\citeauthor{STR:04}) and 36 BBW distances from
\citet{Fouque:etal:03} and \citet{Barnes:etal:03}. The two
calibrations agree to $0.07\mag$ in $V$ at an intermediate period of
$P=10\;$days (\citeauthor{STR:04}). New distances of seven
clusters by \citet{An:etal:07} suggest smaller distances by
$\sim\!0.1$, which brings the two systems to even better
agreement. The BBW distances have been revised twice since 2003
\citep{Gieren:etal:05b,Fouque:etal:07}, but the effect on the
zero point at $P=10\;$days is negligible. The adopted zero point is
$M_{V}=-4.00$. The independent zero point from {\em HST\/} parallaxes
by \citet{Benedict:etal:07} is brighter by 0.05, the one of
\citet{vanLeeuwen:etal:07} by only 0.01.

\subsubsection{The zero point of the P-L relation of LMC}
\label{sec:03:2:2}
The zero point of the LMC P-L relations , which holds for a
metallicity of [O/H]$_{\rm Te}=8.34$ \citep{Sakai:etal:04}, is given
by an {\em adopted\/} distance of LMC. Thirteen determinations from
1997 to 2002 gave a mean modulus of $\mu^{0}=18.54\pm0.02$
(\citeauthor{STR:04}). Sixteen newer determinations are
compiled in Table~\ref{tab:06}. The listed errors of the individual
distance determinations are from the literature, but since they are
incommensurable, a straight mean of $18.53\pm0.01$ has been
derived. The value of $18.54$ is maintained here. Note that none of
the listed distances involves any assumption on the P-L relation,
which would make the calibration circular. 


     If the model P-L relation in $V$ for $Z=0.004, Y=0.25$ of
\citet{Marconi:etal:05} are taken at face value and if they are
combined with the observed Cepheids in LMC one obtains a distance
modulus of $\mu^{0}_{\rm LMC}=18.51\pm0.01$. The result is lower in
$B$ and higher in $I$, because the model colors are still redder than
observed. The distance becomes smaller by $\sim\!0.1\mag$ if the more
realistic model with $Z=0.008, Y=0.25$ are used for LMC.

     An LMC Cepheid at $P=10\;$days is brighter than its Galactic
counterpart by $0.25\mag$. The assumption is devious that this could
be remedied by decreasing the distance of LMC because the zero-point
difference is wavelength-dependent (0.35 in $B$, 0.15 in $I$). The
erroneous assumption of equal zero points has notoriously led to too
small an LMC distance if based on Cepheids.

\subsubsection{The zero point of the P-L relation of SMC}
\label{sec:03:2:3}
The constant terms in equations~(\ref{eq:05})--(\ref{eq:07}) are
calibrated with an adopted SMC modulus of 18.93 (Table~\ref{tab:07})
as mentioned before.    

     In \S~\ref{sec:04} the Cepheid distances shall be compared
with the Pop.~II distance indicators. The calibration of the Cepheids
should therefore be as free of Pop.~II data as possible. In spite of
this, an RR\,Lyr star and a TRGB distance are included for the
zero-point calibration of each of the P-L relations of LMC and SMC
(see Tables~\ref{tab:06} \& \ref{tab:07}). However, their omission
would change the calibration by only $0.02\mag$. In case of LMC such a
change is negligible because the LMC P-L relation is always used in
combination with the independently calibrated Galactic P-L
relation. The SMC P-L relation is used for only three galaxies, which
follow below.


\subsubsection{Metallicity corrections}
\label{sec:03:2:4}
There is a large literature on metallicity corrections to Cepheid
distances (i.e.\ due to differences in the Cepheid P-L relation for
different $Y$ and $Z$ values), and we do not review that literature
here. A fine review is by \citet{Romaniello:etal:05}. For the present
paper we use the formulations by \citeauthor{STT:06}.

     Cepheid distances derived from $V$ and $I$ magnitudes and the
corresponding P-L relations of the {\em Galaxy\/} differ from those
using the P-L relations of LMC. Up to periods of $P\la10{\rm d}$ the
LMC relations yield larger, above this period limit smaller
distances. This was ascribed in \citeauthor{STT:06} to the
metallicity difference of the Cepheids in the 
Galaxy ([O/H]$_{\rm Te}=8.60$) and in LMC ([O/H]$_{\rm Te}=8.34$).
Correspondingly Cepheids with Galactic metallicity are reduced with
the P-L relations of the Galaxy, and those with the LMC metallicity
with the LMC relations. For the distances of Cepheids with
intermediate and slightly extrapolated metallicities an interpolation
formula was derived in \citeauthor{STT:06} (eq.~10).

     The interpolation formula is given here in tabular form for every 
increment of $\Delta$[O/H]$_{\rm Te}=+0.1$  from [O/H]$=8.34$ 
(Table~\ref{tab:08}). The entries give the distance modulus change
$\Delta\mu$ in function of period which must be applied to a distance
derived from $V,I$ photometry and based on the LMC P-L relations. 
The Table can be read with opposite sign if distances from the
Galactic P-L relations are to be corrected to lower metallicities.


For a few galaxies outside the range $8.2<\mbox{[O/H]}_{\rm Te}<8.7$,
the limiting values of 8.2 and 8.7, respectively, have been adopted by
\citeauthor{STT:06}. 

     It may seem paradoxical that metal-rich Cepheids with $\log
P>1.0$ (actually $\log P>0.933$) yield {\em larger\/} distances than
LMC Cepheids although the latter are {\em brighter\/} in $V$ up to
$\log P=1.38$. The reason is that the $V$ and $I$ magnitudes are used
not only to derive a true distance but also the reddening. The
metal-poor Cepheids being blue yield large reddenings leading to large
absorption corrections and hence to small distances. The effect of
metallicity changes on the distance of Cepheids is therefore a
combination of their effect on the luminosities {\em and\/} on the
inferred absorption corrections.

\subsection{A Summary of Available Cepheid Distances}
\label{sec:03:3}
Metallicity-corrected Cepheid distances of 37 galaxies have been
derived in \citeauthor{STT:06} from the Galactic and LMC P-L
relations as given in \citeauthor{STR:04}. 
Six additional Cepheid distances have since become available.

\noindent
NGC\,55. \citet{Pietrzynski:etal:06a} have observed 143 Cepheids in $V$
and $I$ in NGC\,55 with a metallicity of [O/H]$_{\rm Te}=8.35$, i.e.\
close to LMC. Using \citet*{Udalski:etal:99c} LMC P-L relation they
have obtained a modulus of $\mu^{0}=26.45\pm0.05$ if $\mu^{0}_{\rm LMC}$
is at 18.54. From the best 110 Cepheids and the LMC P-L relation of
\citeauthor{STR:04} we obtain $\mu^{0}=26.42$ using
$\langle E(V\!-\!I)\rangle=0.12$ and a small metallicity
correction of $0.01\mag$. If one applies the Galactic P-L relation
instead, which stands for a metallicity of [O/H]$_{\rm Te}=8.60$, one
finds $\mu^{0}(\mbox{Gal})=26.56$ and a metallicity correction of
$-0.16\mag$ from equation~(10) in \citeauthor{STT:06}, resulting
in a corrected modulus of $\mu^{0}=26.40$. We adopt $26.41\pm0.05$ for
NGC\,55.  

\noindent
M\,31 (NGC\,224). \citet{Vilardell:etal:07} have observed hundreds of
badly needed Cepheids in this galaxy, 281 of which the authors
identify as fundamental pulsators. Unfortunately the $B,V$ photometry
of these variables with the 2.5m Isaac Newton Telescope is affected by
blends. \citeauthor{Vilardell:etal:07} have found that Cepheids with
large amplitudes, i.e.\ ${\cal A}_{V}>0.8\mag$, are least blemished by
blends and they have kindly provided to us the subset of the 64 such
fundamental pulsators with $0.4< \log P < 1.6$. Their mean metallicity
is [O/H]$_{\rm Te}=8.66$ from their galactocentric distances and the
metallicity gradients of \citet{Zaritsky:etal:94}. Since this is only
slightly more than the adopted value of Galactic values (8.6), it is
assumed that the M\,31 Cepheids follow the Galactic P-C relation. With
this assumption individual reddenings $E(B\!-\!V)$ were determined,
which turn out to increase with period, the mean value being 
$\langle E(B\!-\!V)\rangle=0.21$. The ensuing absorption-corrected P-L
relations are virtually as {\em steep} ($-2.916\pm0.144$ in $V$) as in
the Galaxy. Comparing these relations with the adopted Galactic P-L
relations yields in $B$ and $V$ $\mu^{0}=24.32\pm0.06$. Had we
compared with LMC at $\mu^{0}=18.54$ the modulus would become $24.18$,
which is still to be increased by $0.11\mag$ for the metallicity
difference to give $\mu^{0}=24.29$ (see \citeauthor{STT:06},
eq.~[10]). However, the distances are still to be corrected for the
amplitude restriction. The largest amplitudes occur in general on the
blue side of the instability strip (\citeauthor{STR:04},
Fig.~11). In the Galaxy the 123 Cepheids with ${\cal A}_{V}>0.8\mag$
from \citet{Berdnikov:etal:00} are bluer in $(B\!-\!V)$ than the total
of 321 Cepheids by $0.02\mag$. If the same value holds for M\,31, the
above reddenings were underestimated by the same amount and the
absorption by $0.06\mag$. The distance becomes then 24.26. On the
other hand blue Cepheids are intrinsically brighter than average
because of the slope $\beta$ of the constant-period lines. Yet, since
$\beta$ is quite flat ($\beta_{V,B-V}=0.6$,
\citeauthor{STR:04}) in the Galaxy and presumably in M\,31
this effect increases the distance by only 0.01, which becomes then
$\mu^{0}=24.27$ for M\,31. -- The Cepheid distance of M\,31 is
significantly smaller than from RR\,Lyr stars (24.53) and the TRGB
(24.47). This may be due to remaining blend effects or to an
overestimate of the reddening, if the metal-rich M\,31 Cepheids are
intrinsically redder than Galactic Cepheids. 
 
\noindent
NGC\,4258. \citet{Macri:etal:06} have observed Cepheids in $B$, $V$,
and $I$ in an outer field of NGC\,4258. They have the same
metallicity  as LMC ([O/H]$_{\rm Te}=8.36$) according to the
metallicity gradient of \citet{Zaritsky:etal:94}. The 36 best Cepheids
in the field yield $\mu^{0}=29.50\pm0.03$ using the LMC P-L relation
of \citeauthor{STR:04} and $\langle E(B\!-\!V)\rangle = 0.042$. 
The Galactic P-L relation yields, after a proper metallicity
correction, the same value. The P-L relation in $B$ shows possibly a
break at $P=10^{\rm d}$, but even if real this has no effect on the
distance determination. For the Cepheids in the inner, metal-rich
field of NGC\,4258 see \S~\ref{sec:03:4:4}.  

\noindent
NGC\,5128 (Cen\,A). Forty-five heavily absorbed Cepheids with $V$ and
$I$ magnitudes from \citet{Ferrarese:etal:07} in the highly peculiar
Galaxy NGC\,5128 yield 
$\langle E(\!V-\!I)\rangle=0.50$, $\mu^{0}=27.62\pm0.04$ and 
$\langle E(\!V-\!I)\rangle=0.42$, $27.71\pm0.04$, respectively,
using the P-L relations of LMC and the Galaxy.  
Since the metallicity of the Cepheids is unknown, a mean of
$\mu^{0}=27.67\pm0.04$ is adopted. Following \citet{Ferrarese:etal:07}
an absorption-to-reddening factor of $R_{V}=2.4$ has been used as
measured for NGC\,5128 by \citet{Hough:etal:87}. Had we assumed the
standard absorption factor of $R_{V}=3.23$ the mean distance would
become $\mu^{0}=27.54$ which is hardly compatible with the TRGB
distance 27.82 \citep{Karachentsev:etal:04} or 27.72
\citep{Rizzi:etal:07}.

     Two more galaxies with known Cepheids have quite low
metallicities, i.e.\ NGC\,3109 and IC\,1613 with [O/H]$_{\rm Te}=8.06$
and 7.86, respectively, from \citet{Sakai:etal:04}, which are close to
SMC ([O/H]$_{\rm Te}=7.98$). In order not to over-extrapolate the
metallicity corrections of \citeauthor{STT:06}, the two galaxies
are tied to the P-L relation of SMC without further metallicity
corrections.

\noindent
NGC 3109. One-hundred-and-two Cepheids from \citet{Pietrzynski:etal:06b}
define, after $2\sigma$ clipping, P-L relations with a slope that is
even flatter than observed in SMC (Table~\ref{tab:05}), but with
the large scatter of $\sigma=0.39$. They indicate, if
compared with SMC, an internal reddening of $E(V\!-\!I)=-0.01\pm0.01$,
which we take as zero,
and a distance modulus of $\mu^{0}=25.41\pm0.04$. 
If the sample is cut at $\log P=0.75$ to
guard against the shortest-period Cepheids being possibly overluminous
in the mean \citep{Sandage:88} yields $\mu^{0}=25.45\pm0.04$, which we
adopt. The small reddening may suggest that the Cepheids are even
bluer than those of SMC. If the restricted sample of Cepheids had been
reduced with the P-L relation of LMC one would have obtained
25.57. \citet{Soszynski:etal:06} have derived $\mu^{0}=25.61$
(if $\mu^{0}_{\rm LMC}=18.54$) from additional magnitudes in $J$ and
$K$ and by comparing with LMC. Earlier work on the Cepheids in 
NGC\,3109 is cited by \citet{Pietrzynski:etal:06b}. 

\noindent
IC\,1613. Forty-two Cepheids from \citet{Antonello:etal:06} fill
exceedingly well the strip in the two-color diagram 
$(B\!-\!V)$ vs. $(V\!-\!I)$ defined by SMC Cepheids. Six
additional Cepheids lie clearly outside that strip and are
omitted. The 42 Cepheids are bluer on average by only
$-0.01\pm0.01\mag$ than SMC Cepheids, which we interprete as zero
reddening. They define P-L relations in $B$, $V$, and $I$ with no
indication of a break and with slopes that are the same within the
errors as the overall slopes in SMC. A comparison of the two sets of
Cepheids yields a distance modulus of $\mu^{0}=24.32\pm0.02$, somewhat
less than $24.50\pm0.12$ from \citeauthor{Antonello:etal:06} who
compared with LMC.

\noindent
WLM may tentatively be compared with the P-L relation of
SMC, although it is very metal-deficient 
([O/H]$_{\rm Te}=7.74$, \citealt{Sakai:etal:04}). 
\citet{Pietrzynski:etal:07} have observed 60 Cepheids, of which three
can be excluded as bright outlyers and one for lack of complete
data. The remaining 56 Cepheids are quite blue and give, if compared
with SMC, $E(V\!-\!I)=-0.03\pm05$, which we take as zero. They define
P-L relations ($\sigma=0.38\mag$) which are even flatter (by
$1\sigma$) than in SMC and significantly flatter than the single-fit
P-L relations of LMC (\citeauthor{STR:04}, eqs. [8\,\&\,9]). No bias
towards bright Cepheids \citep{Sandage:88} is seen at short
periods. Tied to the $V$ and $I$ P-L relations of SMC the Cepheids
give $\mu^{0}=24.80$ and $\mu^{0}=24.83$, respectively. These values
are noticeably smaller than \citet*{Pietrzynski:etal:07} value of
$25.18$ (if LMC at $18.54$), but the adopted value of $\mu^{0}=24.82$
here compares well with the TRGB distance of the galaxy ($24.90$,
Table~\ref{tab:09}), the fit of the entire CMD ($24.88\pm0.09$;
\citealt{Dolphin:00}), and the position of the HB
($24.95\pm0.15$; \citealt{Rejkuba:etal:00}).

\subsection{Are the Metallicity Corrections to Cepheid Distances Reliable?}
\label{sec:03:4}
Any systematic errors of the adopted metallicity corrections must show
by comparing the Cepheid distances with independent distance
indicators. The test is independent of zero-point differences because
we seek only the {\em slope\/} of the function
$\Delta\mu^{0}=f(\mbox{[O/H]})$.

\subsubsection{Comparison of Cepheid distances with TRGB distances}
\label{sec:03:4:1}
Cepheid distances as well as TRGB distances are available for 18
galaxies. The low- and high-metallicity Cepheids in the outer and 
inner field of NGC\,5457 are counted twice (Table~\ref{tab:09}). In
case of NGC\,4258 only the Cepheids in the outer field are plotted for
reasons given in \S~\ref{sec:03:3}. The differences of the respective
distances are plotted against the metallicity [O/H]$_{\rm Te}$ of the
{\em Cepheids\/} in Figure~\ref{fig:05}. The absence of any significant
metallicity dependence on [O/H]$_{\rm Te}$ is striking.   


\subsubsection{Comparison with SNe\,Ia magnitudes}
\label{sec:03:4:2}
Metallicity-corrected Cepheid distances of 10 galaxies were used in
\citeauthor{STS:06} to calibrate the maximum absolute
magnitude of their SNe\,Ia. Any remaining errors
of the metallicity correction will show as an incorrect dependence of
the SN\,Ia luminosities on the Cepheid metallicity [O/H]$_{\rm Te}$. The
test is performed in Figure~\ref{fig:06}. The formal dependence is
statistically insignificant. 


\subsubsection{Comparison of Cepheid distances with velocity distances}
\label{sec:03:4:3}
The difference between the Cepheid distances and velocity distances of
their parent galaxies are not supposed to be a function of the Cepheid
metallicity [O/H]$_{\rm Te}$. The test is difficult because the galaxies
with Cepheid distances are nearby and have small recession velocities
which are substantially influenced by peculiar velocities. Galaxies
with $\mu^{0}<28.2$ are therefore omitted, as are cluster galaxies and
galaxies within $25^{\circ}$ from the Virgo cluster center.  
The distance differences of the remaining 17 galaxies from Table~8 in
\citeauthor{STT:06} and from \S~\ref{sec:03:3} are plotted
against [O/H]$_{\rm Te}$ in Figure~\ref{fig:07}. As in \S~\ref{sec:03:4:1}
the outer- and inner-field Cepheids of NGC\,5457 are plotted
separately and only those of the outer field of NGC\,4258 are
considered. A least-squares-fit to the data results in some dependence
on [O/H]$_{\rm Te}$ (dashed line), but the statistical error is even larger.  

\begin{figure}[t]
   \plotone{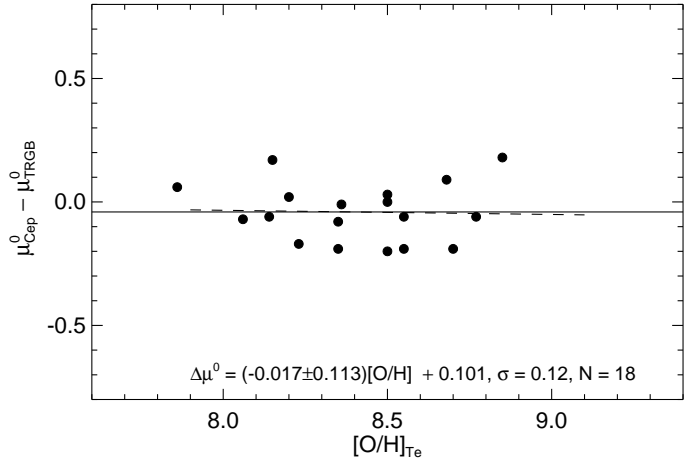}
   \caption{Difference $\Delta\mu^{0}=\mu^{0}_{\rm Cep} -
   \mu^{0}_{\rm TRGB}$ in function of [O/H]$_{\rm Te}$ for the Cepheids.}  
\label{fig:05}
\end{figure}
\begin{figure}[t]
   \plotone{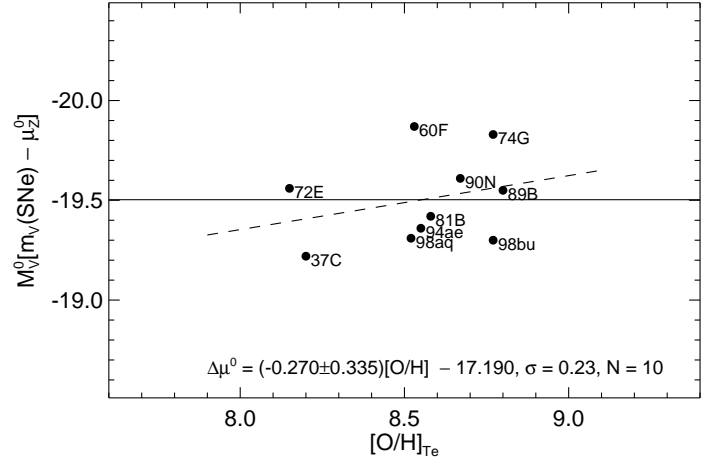}
   \caption{Luminosity of SNe\,Ia in function of the metallicity of
   the Cepheids which led to the distance of the parent galaxy.}
\label{fig:06}
\end{figure}
\begin{figure}[t]
   \plotone{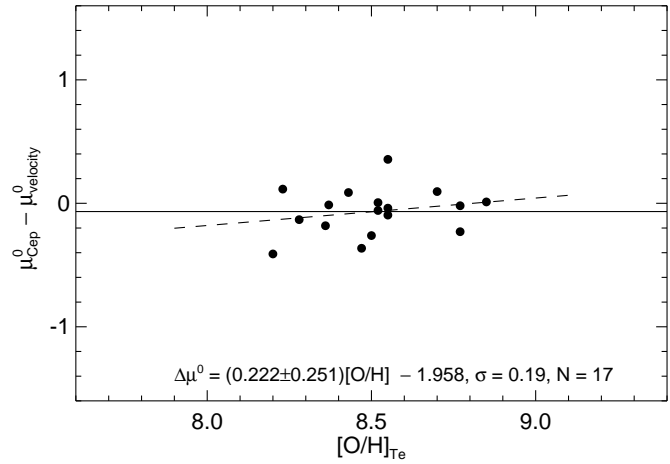}
   \caption{Difference of $\Delta \mu = \mu^{0}_{\rm Cep} -
   \mu^{0}_{\rm velocity}$.}
\label{fig:07}
\end{figure}

If the evidence of Figures~\ref{fig:05}--\ref{fig:07} is combined, the
remaining metal-dependent error of the Cepheid distances amounts to
only $\Delta\mu=(0.05\pm0.10)\,\Delta$[O/H]$_{\rm Te}$. Since $\Delta$[O/H]
is in order of unity the relative distance error between the most
metal-poor and most metal-rich galaxies may in fact be zero, and, in any
case is likely to be $<0.1\mag$. This speaks in favor of the present
metallicity corrections.   

    {\em If\/} all Cepheid distances entering 
Figures~\ref{fig:05}--\ref{fig:07} had been based on the P-L relation
of LMC and if {\em no\/} metallicity corrections had been applied, one
would have found $\Delta\mu=(0.53\pm0.17)\,\Delta$[O/H]$_{\rm Te}$. 
This {\em demonstrates the necessity of metallicity corrections}. 
(It may be noted that the [O/H]$_{\rm Te}$ scale of 
\citet{Kennicutt:etal:03} and \citet{Sakai:etal:04} is compressed by a
factor of $\sim\!1.5$ as compared to the old [O/H] scale which has
been widely used in the present context. Therefore the above relation
translates into $ \Delta\mu\sim 0.35\Delta$[O/H]$_{\rm old}$).

\subsubsection{Comparison with the Cepheids in the inner field of NGC\,4258}
\label{sec:03:4:4}
From the water masers moving on Keplerian orbits about the center of
NGC\,4258 \citet{Herrnstein:etal:99, Herrnstein:etal:05} have derived
a modulus of $29.29\pm0.08\pm0.07$. Both the Cepheid distance
($29.50$; see \S~\ref{sec:03:3}) and TRGB distance ($29.44$; see
\S~\ref{sec:03:3}) are larger, but a mean of $\mu^{0}=29.38$
agrees within $\le0.12\mag$ with all three determinations.   
Cepheids have also been observed in an {\em inner\/} field of
NGC\,4258 by \citet{Macri:etal:06}. Their position in the galaxy with
its chemical gradient \citep{Zaritsky:etal:94} suggest that these
Cepheids are as metal-rich as Galactic Cepheids on average. The
Cepheids are therefore expected to share the P-L relation of the
Galaxy, not the LMC. The 81 Cepheids, complying with the typical
position in the $(B\!-\!V)$ vs. $(V\!-\!I)$ diagram of classical
Cepheids, cover a wide period interval of $0.6<\log P<1.7$. 
Their absorption-corrected P-L relation can be derived by adopting the
above distance of NGC\,4258 using an appropriate P-C relation. 
\citeauthor{Macri:etal:06} have adopted the blue P-C relation of LMC
which, however, is incorrect for the metal-rich and necessarily redder
Cepheids. If one assumes that these Cepheids follow the same P-C
relation {\em as in the Galaxy}, one obtains color excesses
$E(B\!-\!V)$ which are nearly independent of period.
This lends support to the assumption that the P-C relations of the
inner field and of the Galaxy have the same slope. The resulting P-L
relations in $B$, $V$, and $I$ are very flat, in fact in spite of the
high metallicity as flat as in LMC (not considering the break at
$P=10^{\rm d}$) {\em and as flat as in the outer field}. The
observation that the inner-field Cepheids agree with the
LMC Cepheids to within $0.1\mag$ at all periods depends on the
additional {\em assumption\/} that the P-C relations of the inner
field and of the Galaxy do not have only the same slope, but also the
same zero point.

     This is not to suggest that the above combination of a Galactic
P-C relation (for high-metallicity Cepheids) and an LMC P-L relation 
(for low-metallicity Cepheids) could give a consistent
solution. Rather it is likely that the flat slope of the 
{\em metal-rich\/} Cepheids in the inner field of NGC\,4258, in
contrast to Figure~\ref{fig:04}, is caused by a second parameter other
than [O/H], possibly by $Y$ as mentioned before.

     It is fortunate that the P-L relation of the inner field, as
derived here, crosses the Galactic P-L relation at $\log P\sim1.5$,
which happens to be the median period of the known Cepheids in most
galaxies outside the Local Group. It makes therefore little difference
for the derived distances which of the two P-L relations applies to a
given set of high-metallicity Cepheids.

\section{COMPARISON OF THE ZERO POINTS OF THE POP.~I AND POP.~II DISTANCES}
\label{sec:04}
The Cepheid distances of 18 galaxies introduced in 
\S~\ref{sec:03:3} can be compared with their corresponding TRGB
distances (Table~\ref{tab:09}). The comparison is equivalent to a
comparison of Cepheid with RR\,Lyr star distances because the TRGB
distances are so tightly linked with the RR\,Lyr stars through
Table~\ref{tab:02}. On average the Cepheid distances are {\em
  smaller\/} than the TRGB distances by only $0.04\pm0.03$. If instead
the 13 galaxies are compared, for which {\em metal-corrected\/} TRGB
distances are given in the literature, the difference 
$\mu^{0}_{\rm Ceph}-\mu^{0}_{\rm TRGB}$ becomes $-0.02\pm0.03$.
This agreement of the Pop.~I and Pop.~II distances of nearby
galaxies is as good as can possibly be expected.


     \citet{Rizzi:etal:07} have compared the TRGB distances of 15
galaxies with their Cepheid distances, but the latter are derived from
the old P-L relation of \citet{Madore:Freedman:91} without
corrections for metallicity. \citeauthor{Rizzi:etal:07} find
$\langle\mu^{0}_{\rm Ceph}-\mu^{0}_{\rm TRGB} \rangle=0.01\pm0.03$. 
The good agreement is no surprise because it was stated already in
\S~\ref{sec:01} that several old Cepheid distance scales, prior to the
one adopted by \citet{Freedman:etal:01}, agree {\em on average\/} well
with those adopted here.

     Only 10 galaxies used for the comparison by \citet{Rizzi:etal:07}
are also contained in Table~\ref{tab:09}. LMC and SMC are omitted,
because they are used here to calibrate the P-L relation of their
specific metallicity. We do not have a reliable template P-L relation
for the most metal-poor galaxies of \citeauthor{Rizzi:etal:07}
(Sextans A/B).

\section{THE LOCAL AND NOT SO LOCAL HUBBLE DIAGRAMS}
\label{sec:05}
The TRGB and Cepheid distances as well as the Cepheid-calibrated
21cm line width (TF) and SN\,Ia distances can be used to construct
distance-calibrated Hubble diagrams which reach progressively deeper
into the cosmic expansion field.

     The various distances are transformed to the barycenter of the
Local Group which is assumed to lie at the distance of $0.53\;$Mpc in
the direction of M\,31, i.e.\ at two thirds of the way to this galaxy,
because the galaxies outside the Local Group expand presumably away
from the barycenter and not away from the observer. The barycentric
distances are designated with $r^{00}$ and $\mu^{00}$,
respectively. 

     The heliocentric velocities are corrected to the barycenter of
the Local Group following \citet{Yahil:etal:77} and for a
self-consistent Virgocentric infall model assuming a local infall
vector of $220\kms$ and a density profile of the Virgo complex of
$r^{-2}$
\citep{Yahil:etal:80,Dressler:84,Kraan-Korteweg:86,deFreitasPacheco:86,Giraud:90,Jerjen:Tammann:93}. 
The choice of these particular corrections among others proposed in
the literature is justified because they give the smallest scatter in
the Hubble diagrams (see \citeauthor{STS:06}). -- The
velocities are not corrected for the projection angle between the
observer and the Local Group barycenter as seen from the galaxy
because it affects the velocities by less than 2\% for all galaxies
beyond $3\;$Mpc.

\subsection{The Hubble Diagram from TRGB}
\label{sec:05:1}
The galaxies outside the Local Group, for which TRGB distances are
available (\S~\ref{sec:02:2:4}), are plotted in a Hubble diagram
(Fig.~\ref{fig:09}a). 
The nearest galaxies reflect clearly the effect
of the gravitational pull of the Local Group, suggesting that the
zero-velocity surface lies at a distance of $\la\!1.6\;$Mpc from the
barycenter of the Local Group. The 59 galaxies with 
$\mu^{0}_{\rm TRGB}>28.2$ ($4.4\;$Mpc) define a value of
$H_{0}=61.7\pm1.5\ksm$. This is a very local value extending to only
$\mu^{0}=30.4$ ($12\;$Mpc).
The scatter about the Hubble line of $\sigma_{\rm m}=0.39\mag$ cannot
be caused by distance errors, but is explained by (one-dimensional)
peculiar motions of $\sim\!90\kms$ on average
(\citeauthor{STS:06}).

\subsection{The Hubble Diagram from Cepheids}
\label{sec:05:2}
There are 34 galaxies in \citeauthor{STS:06} outside the
Local Group whose Cepheid distances have been derived following the
precepts given in \S\S~\ref{sec:03:1} and \ref{sec:03:2}. Three
additional Cepheid distances are given in \S~\ref{sec:03:3}. The
velocities of the total of 37 galaxies are plotted against $v_{220}$
in Figure~\ref{fig:09}b. The 30 galaxies with 
$\mu^{0}_{\rm Ceph}>28.2$, and excluding the deviating case of
NGC\,3627, define a Hubble line with 
$H_{0}=63.1\pm1.8$ out to $\mu^{0}_{\rm Ceph}\sim 32.0$ ($25\;$Mpc). 
As in the case of the TRGB distances the scatter of
$\sigma_{\rm m}=0.33$ must be caused mainly by peculiar velocities in
the order of $150\kms$ at a median velocity of about $1000\kms$
(assuming a random error of the Cepheid distances of $0.15\mag$).

     The agreement of the local value of $H_{0}$ from TRGB magnitudes
and from Cepheids to within 2\% suggests that the zero-point errors of
the two independent methods do not accumulate to more than
$0.04\mag$.

\subsection{The Hubble Diagram from TF}
\label{sec:05:3}
A complete sample of 104 inclined spiral galaxies with
$v_{220}<1000\kms$ with known 21cm line widths was discussed in
\citeauthor{STS:06}. This small distance limit was
chosen to define as complete a distance-limited sample as
possible. The zero point of the TF distances was calibrated with 31
galaxies for which also Cepheid distances are available from
\citeauthor{STT:06}. The Hubble diagram of the sample galaxies
is repeated in Figure~\ref{fig:09}c from \citeauthor{STS:06},
but added is the mean TF distance of $\mu^{00}=31.61$ of a complete
sample of 49 Virgo cluster spirals plotted at the mean cluster
velocity of $\langle v_{220}\rangle=1152\kms$. Also added is the UMa
cluster with $\mu^{00}=31.45$ and $\langle v_{220}\rangle=1270\kms$.
The TF distance of UMa is taken from \citet{Tully:Pierce:00} who
obtained $\mu^{0}=31.35\pm0.06$ from 38 cluster members with $B,R,I,$
and $K'$ photometry.
After recalibrating their 24 calibrators with the present Cepheid
distances (\citeauthor{STT:06}) one obtains
$\mu^{0}=\mu^{00}=31.45$. The value is adopted here, although the UMa
sample may not be strictly complete as to the faintest cluster spirals.

     The TF distances in Figure~\ref{fig:09}c give a Hubble constant
of $H_{0}=59.0\pm1.9$ out to $\sim\!16\;$Mpc. This agrees well with
$H_{0}$ from TRGB distances (Fig.~\ref{fig:09}a), but it is
$1.6\sigma$ less than determined in Figure~\ref{fig:09}b from
Cepheids. This difference, however, cannot be real because the TF
distances depend entirely on the calibration through Cepheids.
Rather it reflects on the reliability of the TF method. 
In any case the scatter in Figure~\ref{fig:09}c is very large
($0.69\mag$). This 
cannot be attributed to peculiar motions which contribute only
$\sigma_{m}\sim0.3\mag$ in Figures~\ref{fig:09}a \& b. Even if
some of the scatter is caused by observational errors of the input
parameters, the intrinsic dispersion is large. This makes the TF
method vulnerable to Malmquist bias if magnitude-limited samples are
used instead of complete distance-limited samples. This is the reason
why the more distant clusters of \citet{Tully:Pierce:00}, which are
expected to suffer at least some magnitude bias, are not considered
here. 

\begin{figure}[t]
   \plotone{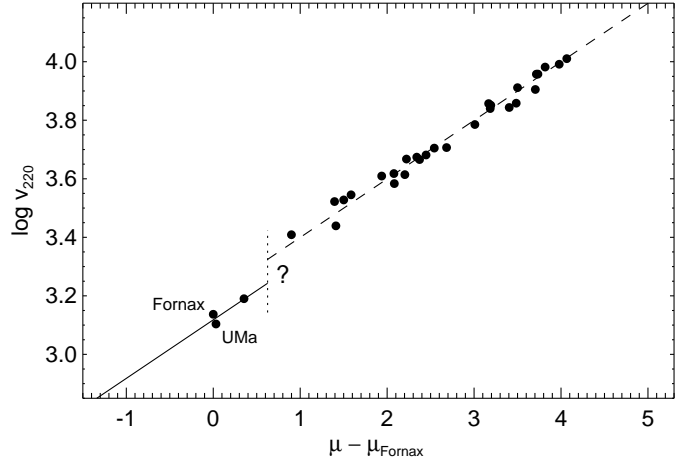}
   \caption{Hubble diagram of 32 clusters with {\em relative\/}
   21cm line width distances from \citet{Masters:etal:06}. The
   zero point is arbitrarily set at the Fornax cluster. Note the
   spurious break of the Hubble line (see text).}
\label{fig:08}
\end{figure}

\begin{figure}[t]
   \epsscale{0.75}
   \plotone{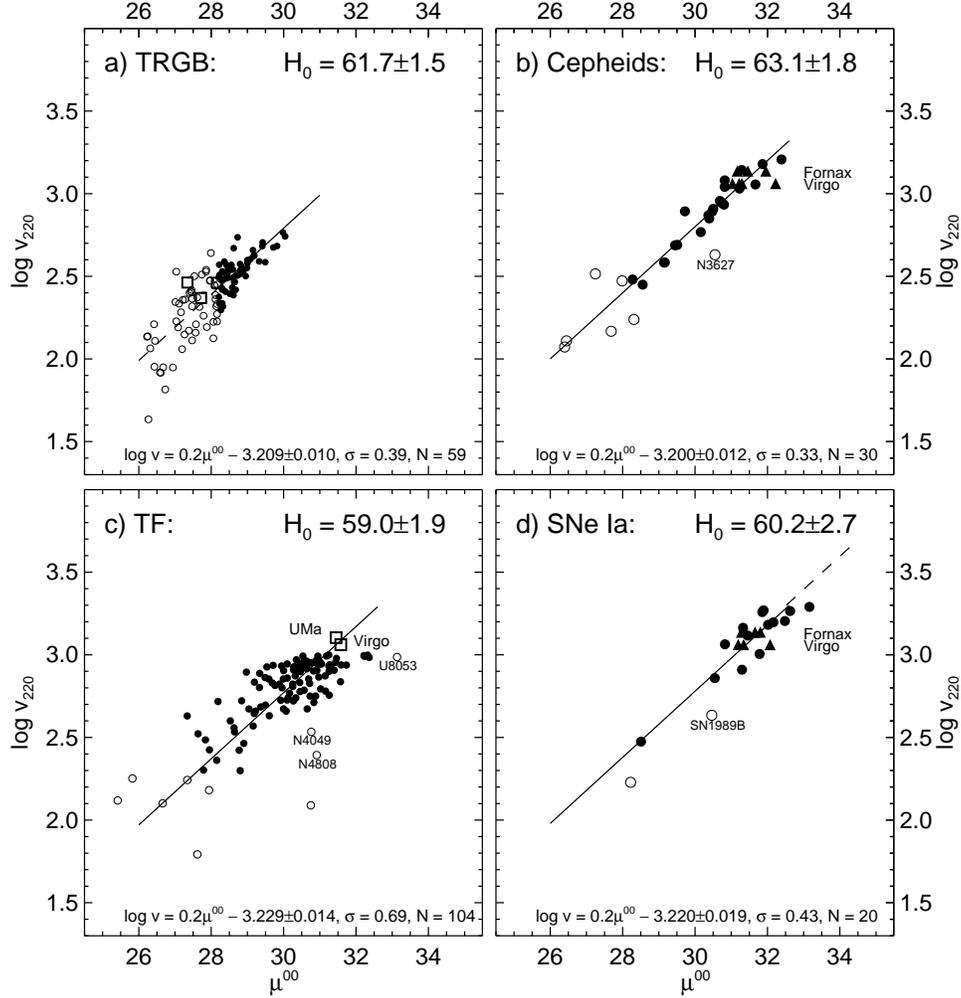}
   \caption{Distance-calibrated Hubble diagrams for a) TRGB distances;
   the M\,81, Cen\,A, and IC\,342 groups are shown as squares at their
   mean position;
   b) Cepheid distances (the Virgo und Fornax cluster members are
   plotted at $v_{220}=1152$ and $1371\kms$, respectively); c)
   21cm line width distances of a complete sample of field galaxies 
   with $v_{220}<1000\kms$; the Virgo cluster and the UMa cluster (at 
   $v_{220}=1236\kms$) are also shown; d) SN\,Ia distances with 
   $v_{220}<2000\kms$; the dashed line is the downwards extension of
   the Hubble line defined by 62 SNe\,Ia with $3000<v_{\rm
   CMB}<20,000\kms$ and reflecting the large-scale value of $H_{0}$
   (from \citeauthor{STS:06}). Triangles denote cluster
   members. Open symbols are objects with 
   $\mu^{0}<28.2$ or in c) with $v_{220}<200\kms$ and a few deviating 
   objects (identified); open symbols are not considered for the 
   solution.}
\label{fig:09}
\end{figure}

     \citet{Masters:etal:06} have measured $I$-band TF distances for
an average of 25 galaxies in 31 clusters between 
$1100<v_{\rm CMB}\la10,000\kms$. The clusters define an impressively
tight Hubble diagram with a scatter of $\sigma_{m}\approx0.15\mag$,
comparable only to distant SNe\,Ia
(\citeauthor{RTS:05}). However the diagram has no zero-point
calibration and does not per se define a value of $H_{0}$. The authors
propose to calibrate their TF relation by local galaxies with Cepheid
distances. Yet, however fair their selection criteria may be for the
galaxies in the different clusters, the same criteria cannot be
applied to a distance-limited, yet highly incomplete sample of field
galaxies with Cepheid distances.  
This would be decisive in view of the large intrinsic scatter of the
TF relation. Therefore any value of $H_{0}$ derived from the two sets
of differently selected galaxies remains unreliable. 
A safer way would be to calibrate the Hubble diagram with the nearest
two clusters of their sample with independently known distances, i.e.\
the Fornax and UMa clusters. However, as seen in Figure~\ref{fig:08}
their relative distances do not put the nearest clusters with
$v_{220}<2000\kms$ on the same Hubble line as the more distant
clusters. The latter are shifted by $\Delta\log v =0.08$ 
or $\Delta\mu =0.40$, as compared to the nearest clusters. The
corresponding increase of $H_{0}$ by $\sim\!20\%$ at $\sim2000\kms$ is
denied by the Hubble diagram of SNe\,Ia (see e.g.\ Fig.~\ref{fig:09}d;
\citeauthor{RTS:05}, Fig.~15). The break of the Hubble line
suggests that the selection criteria for the individual galaxies in
the near and distant clusters resulted in two incompatible cluster
samples. The unrealistic break of the Hubble line is not caused by
corrections of the velocities for the CMB motion. It persists whether
the nearest clusters are corrected for CMB motion or not. In
Figure~\ref{fig:08} the nearest clusters are {\em not\/} corrected for
this motion because the co-moving volume extends to at least
$3000\kms$ (\citealt{Federspiel:etal:94}, Figs. 17--19;
\citealt{Dale:Giovanelli:00}). 


\subsection{The Hubble Diagram from SN\,Ia Distances}
\label{sec:05:4}
Figure~\ref{fig:09}d shows the Hubble diagram of local SNe\,Ia with
$v_{220}<2000\kms$. The SNe\,Ia are drawn from the homogeneously
reduced list of SN\,Ia magnitudes (\citeauthor{RTS:05}). Their
mean Cepheid-calibrated absolute magnitude is adopted as
$M^{0}_{V}(\max)=-19.46$ (\citeauthor{STS:06}). Of the 22
SNe\,Ia, one has $\mu^{0}<28.2$ and SN\,1989B in NGC\,3627 is an
outlyer. Three SNe\,Ia each in the Virgo and Fornax cluster are
plotted at their mean cluster velocity. The 20 adopted SNe\,Ia give
$H_{0}=60.2\pm2.7$ with a scatter of $\sigma_{m}=0.43$. Both
values are statistical the same as those derived from Cepheids in
Figure~\ref{fig:09}b. The statistical agreement in $H_{0}$ must be
expected because the zero point of the SNe\,Ia depends entirely on the
Cepheids, but the SNe\,Ia extend the Hubble diagram to $30\;$Mpc and
beyond (see below). The similar scatter of the SNe\,Ia and Cepheids in 
their respective Hubble diagrams suggests that they are equally
good distance indicators.

     The weighted mean of $H_{0}$ from TRGB distances, Cepheids, TF
distances, and SNe\,Ia is $61.3\pm1.0$.
There is no hint that the mean value of $H_{0}$ varies significantly
from about $4$ to $30\;$Mpc. Clear deviations from a steady Hubble
flow are detected only from the pull of the Local Group and from the
Virgocentric flow. Other deviations near local mass concentrations are
expected to exist \citep[e.g.][their Figs. 5--7]{Klypin:etal:03}, but
the present method considering relatively few 
galaxies is not suitable to detect them. The distance independence of
the {\em mean\/} value of $H_{0}$, however, is the more significant as
the distant SNe\,Ia with $3000<v_{\rm CMB}<20,000\kms$ yield the same
value of $H_{0}=62.3$. In spite of all mass clusterings the overall
value of $H_{0}$ does not depend on distance.


\section{CONCLUSIONS AND OUTLOOK}
\label{sec:06}
The local agreement of the Pop.~I and Pop.~II distance scales is
encouraging. Pop.~I Cepheids as well as SNe\,Ia and the TF relation,
both calibrated through Cepheids, on the one hand and Pop.~II
RR\,Lyr stars and the magnitude of the TRGB, based on these stars,
on the other, yield highly consistent distances for many
individual galaxies. 
Moreover they all agree with a local value of the Hubble constant of
$H_{0}=62.3$. Finally the SNe\,Ia carry the distance scale into the
cosmic expansion field out to $\sim\!20,000\kms$ and prove that
$H_{0}$ is virtually unchanged in the free field.  

     The Pop.~II distance scale alone leads through RR\,Lyr stars and
the TRGB to a {\em minimum\/} distance of the Virgo cluster of
$\mu^{0}\ge31.3$. If one wants to drive the Pop.~II distances to
cosmic scales, one may note that the four SNe\,Ia discussed in
\S~\ref{sec:02:2:5} give a preliminary mean TRGB-calibrated luminosity
of  $M_{V}=-19.37\pm0.06$. Yet the value of $M_{V}=-19.46\pm0.04$ from
10 Cepheid-calibrated SNe\,Ia (\citeauthor{STS:06}) has still much
higher weight. Nevertheless the agreement is encouraging at this
stage. Other TRGB-calibrated SNe\,Ia will become available in the
future.    

     Even if the agreement of $H_{0}$ from independent distances of
Population~I and II objects is accidental, it is now unlikely that the
systematic error of $H_{0}$ is as large as 10\%. If the systematic
error is in fact as high as $0.2\mag$, or 10\% in distance, this
translates to $\pm5$ units in $H_{0}$, as stated in the Abstract. 

     If the value of $H_{0}=62$ is taken at face value and combined
with WMAP data of the CMB fluctuation spectrum it poses constraints on
the equation of state $w=p/\rho$ of the dark energy. According to
\citet{Spergel:etal:07} the WMAP3 data give 
$\Omega_{m}h^{2}=0.128\pm0.008$, ($h\equiv H_{0}/100$), 
from which follows then a rather high matter density parameter of
$\Omega_{m}=0.33$. This value disfavors a Universe with $w=-1$ at
the $2\sigma$ level \citep[see Figs.~15 and 16 of][]{Spergel:etal:07}
and suggests a quintessence model with $w>-1$.
The high matter density $\Omega_{m}$ is not favored, however, by
the large-scale distribution of the luminous red galaxies in the Sloan
Digital Sky Survey if it is combined with the WMAP3 data. In that case
a closed Universe with $\Omega_{\rm total}\sim1.02$ is compatible with
$H_{0}=62$ \citep[][Fig.~13]{Tegmark:etal:06}. This illustrates that a
reliable value of $H_{0}$ imposes stringent constraints on any
cosmological models.

\acknowledgments
We thank Dres.\ Abhijit Saha and Norbert Straumann for helpful
discussions. 
Dr. Francesc Vilardell has kindly made available the selected sample
of Cepheids in M\,31.
A.\,S. thanks the Observatories of the Carnegie Institution for
post-retirement facilities.
We thankfully acknowledge the helpful comments of an anonymous
referee.



\setlength\textheight{9.0in}%

\clearpage


\begin{deluxetable}{lcclccccclclc}
\tablewidth{0pt}
\tabletypesize{\footnotesize}
\tablecaption{RR Lyr star distances of 24 galaxies (in order of RA).\label{tab:01}}   
\tablehead{
 \colhead{Name}               & 
 \colhead{$N_{\rm RR}$}       &
 \colhead{[Fe/H]}             &
 \colhead{$\langle V\rangle$} &
 \colhead{$E(B\!-\!V)$}       &
 \colhead{$A_{V}$}            &
 \colhead{$V^{0}$}            &
 \colhead{$M^{V}_{\rm San}$}  &
 \colhead{$M^{V}_{\rm Lit}$}  &
 \colhead{$\mu^0_{\rm new}$}  &
 \colhead{$\mu^0_{\rm Lit}$}  & 
 \colhead{Tel.}               &
 \colhead{Ref}          
\\
 \colhead{(1)}     & 
 \colhead{(2)}     & 
 \colhead{(3)}     & 
 \colhead{(4)}     & 
 \colhead{(5)}     & 
 \colhead{(6)}     & 
 \colhead{(7)}     & 
 \colhead{(8)}     & 
 \colhead{(9)}     & 
 \colhead{(10)}    & 
 \colhead{(11)}    & 
 \colhead{(12)}    & 
 \colhead{(13)}    
} 
\startdata
NGC\,147   &  36 & -1.37 & 25.29 & 0.173 & 0.54 & 24.75 & 0.55 & 0.77    & 24.20 & 23.92   & 200'' & 1   \\
And III    &  39 & -1.88 & 25.01 & 0.057 & 0.18 & 24.83 & 0.47 & 0.50    & 24.36 & 24.33   & HST   & 2   \\
NGC\,185   & 151 & -1.37 & 25.24 & 0.182 & 0.56 & 24.68 & 0.55 & 0.77    & 24.13 & 23.79   & 200'' & 3   \\
NGC\,205   &  30 & -0.85 & 25.54 & 0.062 & 0.19 & 25.35 & 0.70 & 0.77    & 24.65 & 24.65   & 200'' & 4   \\
NGC\,224   &  54 & -1.60 & 25.30 & 0.062 & 0.19 & 25.11 & 0.51 & 0.55    & 24.60 & 24.55   & HST   & 5   \\
And I      &  72 & -1.46 & 25.14 & 0.054 & 0.17 & 24.97 & 0.53 & 0.57    & 24.44 & 24.40   & HST   & 6   \\
SMC        & 514 & -1.70 & 19.74 & 0.087 & 0.27 & 19.47 & 0.49 & \nodata & 18.98 & \nodata & 1.3m  & 7   \\
Sculptor   & 226 & -1.70 & 20.14 & 0.018 & 0.06 & 20.08 & 0.49 & 0.43    & 19.59 & 19.65   & 40''  & 8   \\
IC\,1613   &  13 & -1.30 & 25.00 & 0.025 & 0.08 & 24.92 & 0.57 & 0.60    & 24.35 & 24.32   & HST   & 9   \\
And II     &  72 & -1.49 & 24.87 & 0.062 & 0.19 & 24.68 & 0.53 & 0.57    & 24.15 & 24.06   & HST   & 10  \\
NGC\,598   &  43 & -1.30 & 25.12 & var   & var  & 25.34 & 0.57 & 0.67    & 24.77 & 24.67   & HST   & 11  \\
Phoenix    &   4 & -1.40 & 23.64: & 0.016 & 0.05 & 23.59 & 0.54 & \nodata & 23.05: & \nodata & 4m  & 12  \\
Fornax     & 197 & -1.95 & 21.27 & 0.042 & 0.13 & 21.14 & 0.47 & 0.48    & 20.67 & 20.66   & HST   & 13a \\
Fornax     &  72 & -1.78 & 21.28 & 0.042 & 0.13 & 21.15 & 0.48 & 0.44    & 20.67 & 20.72   & 6.5m  & 13b \\
Fornax     & 500 & -1.81 & 21.38 & 0.042 & 0.13 & 21.25 & 0.48 & \nodata & 20.77 & 20.75   & 1.3m  & 13c \\
LMC        & 108 & -1.46 & 19.37 & 0.101 & 0.31 & 19.06 & 0.53 & 0.61    & 18.53 & 18.45   & 1.54m & 14  \\
Carina     &  58 & -1.90 & 20.76 & 0.063 & 0.20 & 20.56 & 0.47 & 0.58    & 20.09 & 20.10   & 4m    & 15a \\
Carina     &  33 & -2.2: & 20.69 & 0.063 & 0.20 & 20.49 & 0.40 & 0.57    & 20.09 & 19.93   & 1.3m  & 15b \\
Leo A      &   8 & -1.70 & 25.10 & 0.021 & 0.07 & 25.03 & 0.49 & 0.53    & 24.54 & 24.51   & 3.8m  & 16  \\
Leo I      &  74 & -1.82 & 22.60 & 0.036 & 0.11 & 22.49 & 0.48 & 0.44    & 22.01 & 22.04   & 2.2m  & 17  \\
Sextans    &  36 & -1.60 & 20.36 & 0.050 & 0.16 & 20.20 & 0.51 & 0.57    & 19.69 & 19.67   & 1m    & 18  \\
Leo II     &  80 & -1.90 & 22.10 & 0.017 & 0.05 & 22.05 & 0.47 & 0.44    & 21.58 & 21.66   & 3.6m  & 19  \\
UMi        &  82 & -1.90 & 19.86 & 0.032 & 0.10 & 19.76 & 0.47 & 0.60    & 19.29 & 19.35   & 3.52m & 20  \\
Draco      &  94 & -1.60 & 20.18 & 0.027 & 0.08 & 20.10 & 0.51 & 0.69    & 19.59 & 19.61   & 1.2m  & 21  \\
Sag dSph   &  63 & -1.79 & 18.17 & 0.153 & 0.47 & 17.70 & 0.48 & 0.52    & 17.22 & 17.19   & 0.9m  & 22  \\
NGC\,6822  &  15 & -1.92 & 24.63 & 0.236 & 0.73 & 23.90 & 0.47 & 0.50    & 23.43 & 23.41   & VLT   & 23  \\
And VI     &  91 & -1.58 & 25.30 & 0.064 & 0.20 & 25.10 & 0.51 & 0.55    & 24.59 & 24.56   & 2.5m  & 24  \\
\enddata
\tablenotetext{\,}{References. --- 
 (1)  \citealt{Saha:etal:90};
 (2)  \citealt{Pritzl:etal:05};
 (3)  \citealt{Saha:Hoessel:90};
 (4)  \citealt{Saha:etal:92};
 (5)  \citealt{Brown:etal:04};
 (6)  \citealt{Pritzl:etal:05};
 (7)  \citealt{Soszynski:etal:02};
 (8)  \citealt{Kaluzny:etal:95};
 (9)  \citealt{Dolphin:etal:01};
 (10) \citealt{Pritzl:etal:04};
 (11) \citealt{Sarajedini:etal:06};
 (12) \citealt{Gallart:etal:04};
(13a) \citealt{Mackey:Gilmore:03};
(13b) \citealt{Greco:etal:05};
(13c) \citealt{Bersier:Wood:02};
 (14) \citealt{Clementini:etal:03a,Soszynski:etal:03,Alcock:etal:04,Borissova:etal:04};
(15a) \citealt{Saha:etal:86};
(15b) \citealt{Udalski:00};
 (16) \citealt{Dolphin:etal:02};
 (17) \citealt{Held:etal:00,Held:etal:01}; 
 (18) \citealt{Mateo:etal:95a};
 (19) \citealt{Demers:Irwin:93,Siegel:Majewski:00};
 (20) \citealt{Nemec:etal:88,Bellazzini:etal:02,Carrera:etal:02};
 (21) \citealt{Bonanos:etal:04,Grillmair:etal:98,Nemec:85,Aparicio:etal:01};
 (22) \citealt{Layden:Sarajedini:00,Mateo:etal:95b};
 (23) \citealt{Clementini:etal:03b,McAlary:etal:83};
 (24) \citealt{Pritzl:etal:02}.
}
\end{deluxetable}
\setlength\textheight{8.4in}%

\begin{deluxetable}{lcclcc}
\tablewidth{0pt}
\tabletypesize{\footnotesize}
\tablecaption{Calibration of the TRGB by means of RR Lyr stars.\label{tab:02}}   
\tablehead{
 \colhead{Name}   & 
 \colhead{$(V\!-\!I)^{\rm TRGB}$}  &
 \colhead{$\mu^{0}_{\rm RR}$} &    
 \colhead{$m^{\rm TRGB}_{I}$} &    
 \colhead{$M^{\rm TRGB}_{I}$} &    
 \colhead{Ref}
\\
 \colhead{(1)}     & 
 \colhead{(2)}     & 
 \colhead{(3)}     &
 \colhead{(4)}     & 
 \colhead{(5)}     & 
 \colhead{(6)}     
} 
\startdata
Leo A     & 1.33    & 24.54 & 20.53        &  -4.01  & 1      \\
Sex dSph  & 1.35    & 19.69 & 15.78        &  -3.91  & 2,3    \\
And I     & 1.40    & 24.44 & 20.49        &  -3.95  & 4      \\
UMi       & 1.40    & 19.29 & 15.20        &  -4.09  & 5      \\
SMC       & 1.45    & 18.98 & 14.95        &  -4.03  & 6      \\
Sculptor  & 1.47    & 19.59 & 15.57        &  -4.02  & 7      \\
Draco     & 1.48    & 19.59 & 15.62        &  -3.97  & 5      \\
And II    & 1.51    & 24.15 & 20.11        &  -4.04  & 4      \\
Carina    & 1.54    & 20.09 & 16.03        &  -4.06  & 8      \\
Leo I     & 1.55    & 22.01 & 17.95        &  -4.06  & 9      \\
IC\,1613  & 1.56    & 24.35 & 20.24        &  -4.11  & 7      \\
Leo II    & 1.60    & 21.58 & 17.56        &  -4.02  & 2      \\
Phoenix   & 1.60    & 23.05: & 19.17       & (-3.88) & 2      \\
Fornax    & 1.61    & 20.67 & 16.68        &  -3.99  & 10     \\
NGC\,598  & 1.65    & 24.77 & 20.65        &  -4.12  & 4,7,11 \\
NGC\,6822 & 1.65    & 23.43 & 19.35        &  -4.08  & 12     \\
And III   & 1.69    & 24.36 & 20.35        &  -4.01  & 13     \\
LMC       & 1.70    & 18.53 & 14.54        &  -3.99  & 7      \\
NGC\,147  & 1.70    & 24.20 & 20.20        &  -4.00  & 13,14  \\
And VI    & 1.71    & 24.59 & 20.45        &  -4.14  & 13     \\
NGC\,205  & 1.71    & 24.65 & 20.53        &  -4.12  & 13     \\
NGC\,185  & 1.76    & 24.13 & 19.98        &  -4.15  & 7,13   \\
NGC\,224  & 1.89    & 24.60 & 20.46        &  -4.14  & 7,13   \\
Sag dSph  & \nodata & 17.22 & 12.46$^{1)}$ & (-4.76) & 2      \\
\tableline
{\bf mean} &       &       &       &
           \multicolumn{2}{l}{\boldmath{$-4.05\pm0.02$}}\\
           &       &       &       &
           \multicolumn{2}{l}{$\sigma=0.08, N=22$}

\enddata
\tablenotetext{1)}{Poorly defined} 
\tablerefs{References to $m^{\rm TRGB}_{I}$:
 (1) \citealt{Dolphin:etal:03};
 (2) \citealt{Karachentsev:etal:04};
 (3) \citealt{Lee:etal:03};
 (4) \citealt{McConnachie:etal:04};
 (5) \citealt{Bellazzini:etal:02};
 (6) \citealt{Udalski:00}; \citealt{Cioni:etal:00};
 (7) \citealt{Rizzi:etal:07};
 (8) \citealt{Smecker-Hane:etal:94,Udalski:98};
 (9) \citealt{Bellazzini:etal:04b};
 (10) \citealt{Bersier:00};
 (11) \citealt{Galleti:etal:04};
 (12) \citealt{Gallart:etal:96};
 (13) \citealt{McConnachie:etal:05};
 (14) \citealt{Han:etal:97}.
}
\end{deluxetable}

\begin{deluxetable}{lrlclcccl}
\tablewidth{0pt}
\tabletypesize{\footnotesize}
\tablecaption{A tentative TRGB calibration of the SN\,Ia luminosity.\label{tab:03}}
\tablehead{
 \colhead{SN}   & 
 \colhead{$m_{V}^{0}$} &
 \colhead{galaxy} &
 \colhead{$\mu_{\rm TRGB}^{0}$} &
 \colhead{group} &
 \colhead{$\langle \mu_{\rm TRGB}^{0}\rangle$} &
 \colhead{n} &
 \colhead{Ref.} &
 \colhead{$M_{V}^{0}$(SN\,Ia)} 
\\
 \colhead{(1)}     & 
 \colhead{(2)}     & 
 \colhead{(3)}     &
 \colhead{(4)}     & 
 \colhead{(5)}     & 
 \colhead{(6)}     &  
 \colhead{(7)}     & 
 \colhead{(8)}     & 
 \colhead{(9)}     
}
\startdata
1937C  &  8.99 &  IC\,4182 &  28.21  & CnV~II & 28.26 & 16 & 1 & $-19.22$ \\ 
1972E  &  8.49 & NGC\,5253 &  27.89  & Cen\,A & 27.89 & 24 & 2 & $-19.40$ \\
1989B  & 10.95 & NGC\,3627 & \nodata & Leo~I  & 30.43 & 2  & 3 & $-19.48$ \\
1998bu & 11.04 & NGC\,3368 & \nodata & Leo~I  & 30.43 & 2  & 3 & $-19.39$ \\ 
\tableline
\multicolumn{8}{l}{\bf mean}                 & \boldmath{$-19.37\pm0.06$}
\enddata
%
\vspace*{-0.4cm}                    
\tablerefs{
 (1)  \citealt{Sakai:etal:04}; \citealt{Rizzi:etal:07}, 
 (2)  \citealt{Sakai:etal:04},
 (3)  \citealt{Sakai:etal:04} for NGC\,3351; \citealt{Sakai:etal:97}
 for NGC\,3379.
}
\end{deluxetable}

\begin{deluxetable}{cclcclcc}
\tablewidth{0pt}
\tabletypesize{\footnotesize}
\tablecaption{Slopes of P-C relations in Galaxy, LMC, and SMC. (Fits
  for $0.4\le \mbox{log} P \le 1.0$)\label{tab:04}}
\tablehead{
 \colhead{}         & 
 \colhead{Galaxy}   &
 \colhead{}         & 
 \colhead{LMC}      &
 \colhead{$\Delta$} &
 \colhead{}         & 
 \colhead{SMC}      &
 \colhead{$\Delta$}
\\
\cline{4-5}
\cline{7-8}
 \colhead{(1)}      & 
 \colhead{(2)}      & 
 \colhead{}         & 
 \colhead{(3)}      &
 \colhead{(4)}      & 
 \colhead{}         & 
 \colhead{(5)}      & 
 \colhead{(6)}
}
\startdata
$(B\!-\!V)^{0}$ & $0.366\pm0.15$ && $0.273\pm0.024$ & $0.093\pm0.028$ &
                                  & $0.198\pm0.024$ & $0.168\pm0.028$ \\  
$(V\!-\!I)^{0}$ & $0.256\pm0.15$ && $0.160\pm0.022$ & $0.096\pm0.027$ &
                                  & $0.199\pm0.024$ & $0.057\pm0.027$ \\  
\enddata
%
\end{deluxetable}

\clearpage

\begin{deluxetable}{lcccccl}
\tablewidth{0pt}
\tabletypesize{\footnotesize}
\tablecaption{Metallicities and P-L slopes of nine galaxies.\label{tab:05}}   
\tablehead{
 \colhead{Galaxy}   & 
 \colhead{[O/H]$_{\rm Te}$} &
 \colhead{slope $B$} &
 \colhead{slope $V$} &
 \colhead{slope $I$} &
 \colhead{error $V$} &
 \colhead{original source}      
} 
\startdata
NGC\,3351        & 8.85 & \nodata & $-3.12$ & $-3.38$ & 0.39 & \citealt{Graham:etal:97} \\
NGC\,4321        & 8.74 & \nodata & $-3.17$ & $-3.43$ & 0.34 & \citealt{Ferrarese:etal:96} \\
M\,31            & 8.66 & $-2.55$ & $-2.92$ & \nodata & 0.21 & \citealt{Vilardell:etal:07} \\
Galaxy           & 8.60 & $-2.69$ & $-3.09$ & $-3.35$ & 0.09 & \citeauthor{STR:04} \\
LMC\tablenotemark{1)}      & 8.34 & $-2.34$ & $-2.70$ & $-2.94$ & 0.03 & \citealt{Udalski:etal:99a} \\
NGC\,6822\tablenotemark{2)} & 8.14 & \nodata & $-2.49$ & $-2.81$ & 0.10 & \citealt{Pietrzynski:etal:04} \\ 
NGC\,3109        & 8.06 & \nodata & $-2.13$ & $-2.40$ & 0.18 & \citealt{Pietrzynski:etal:06b} \\ 
SMC$^{3)}$       & 7.98 & $-2.22$ & $-2.59$ & $-2.86$ & 0.05 & \citealt{Udalski:etal:99b} \\
IC\,1613         & 7.86 & $-2.36$ & $-2.67$ & $-2.80$ & 0.12 & \citealt{Antonello:etal:06} \\ 
WLM              & 7.74 & \nodata & $-2.52$ & $-2.74$ & 0.15 & \citealt{Pietrzynski:etal:07} \\ 
Sextans A+B      & 7.52 & $-1.43$ & $-1.59$ & $-1.47$ & 0.39 & \citealt{Piotto:etal:94} 
\enddata
%
\vspace*{-0.3cm}                    
{\tablenotetext{1)}{Single-fit slope, neglecting the break at
  $P=10^{\rm d}$.}
\tablenotetext{2)}{Because of large scatter the slope depends
  somewhat on the period cut-off; here $P\ge5.5^{\rm d}$.}
\tablenotetext{3)}{Omitting Cepheids {\em below\/} $P=2.5^{\rm
  d}$.}}
\end{deluxetable}

\begin{deluxetable}{lcl}
\tablewidth{0pt}
\tabletypesize{\footnotesize}
\tablecaption{Distance of LMC from literature since 2002.\label{tab:06}}   
\tablehead{
 \colhead{Author(s)}   & 
 \colhead{$(m-M)^{0}$} &
 \colhead{Method}      
} 
\startdata
\citealt{Fitzpatrick:etal:02}    & $18.50\pm0.05$ & eclipsing binary HV982\\
\citealt{Fouque:etal:03}         & $18.55\pm0.04$ & BBW\\
\citealt{Clausen:etal:03}        & $18.63\pm0.08$ & eclipsing binaries\\
\citealt{Clementini:etal:03a}    & $18.52\pm0.09$ & review\\
\citealt{Groenewegen:Salaris:03} & $18.58\pm0.08$ & main sequence of NGC1866\\
\citealt{Salaris:etal:03}        & $18.47\pm0.01$ & red-clump stars\\
\citealt{Storm:etal:04}          & $18.48\pm0.07$ & BBW\\
\citealt{Feast:04}               & $18.48\pm0.08$ & Miras\\
\citealt{Dall'Ora:etal:04}       & $18.52\pm0.03$ & semi-theoretical\\
\citealt{Sakai:etal:04}          & $18.59\pm0.09$ & TRGB\\
\citealt{Alves:04}               & $18.50\pm0.02$ & review\\
\citealt{Panagia:05}             & $18.56\pm0.05$ & SN1987A light echo\\
\citealt{Gieren:etal:05b}        & $18.53\pm0.06$ & BBW\\
\citealt*{Sandage:Tammann:06}    & $18.55\pm0.10$ & RR Lyr\\
\citealt{Keller:Wood:06}         & $18.54\pm0.02$ & bump Cepheids\\
\citealt{Sollima:etal:06}        & $18.54\pm0.15$ & RR Lyr in $K$ band\\
\tableline
{\bf mean}       & \boldmath{$18.53_{4}\pm0.01_{1}$} & 
\enddata
\end{deluxetable}

\begin{deluxetable}{lcl}
\tablewidth{0pt}
\tabletypesize{\footnotesize}
\tablecaption{The distance of SMC from literature since 2004.\label{tab:07}}   
\tablehead{
 \colhead{Author(s)}   & 
 \colhead{$(m-M)^{0}$} &
 \colhead{Method}      
} 
\startdata
\citealt{Sakai:etal:04}          & $18.96\pm0.10$ & TRGB\\
\citealt{Storm:etal:04}          & $18.88\pm0.14$ & BBW\\
\citealt{Hilditch:etal:05}       & $18.91\pm0.03$ & eclipsing binary\\
\citealt*{Sandage:Tammann:06}    & $18.96\pm0.10$ & RR Lyr\\
\citealt{Keller:Wood:06}         & $18.93\pm0.02$ & bump Cepheids\\
\tableline
{\bf mean}             & \boldmath{$18.93\pm0.02$} & 
\enddata
\end{deluxetable}

\begin{deluxetable}{clccccc}
\tablewidth{0pt}
\tabletypesize{\footnotesize}
\tablecaption{Distance modulus corrections to be applied to distances
  derived from the LMC P-L relations in $V$ and $I$ for any increase
  of the metallicity by $\Delta\mbox{[O/H]}=0.1$ from [O/H]$^{\rm
    LMC}_{\rm Te}=8.34$.\label{tab:08}} 
 \tablehead{$\log P$    & & $0.50$ & $0.75$ & $1.00$ & $1.25$ & $1.50$}
\startdata
$\Delta\mu$ & & $-0.07$ & $-0.03$ & $+0.01$ & $+0.05$ & $+0.09$ 
\enddata
\end{deluxetable}

\setlength\textheight{9.2in}%
\clearpage

\begin{deluxetable}{lccccccc}
\tablewidth{0pt}
\tabletypesize{\footnotesize}
\tablecaption{Comparison of the \citeauthor{STT:06} Cepheid
  distances$^{1)}$ with TRGB distances.\label{tab:09}}   
\tablehead{
 \colhead{Name}                 & 
 \colhead{$\mu^{0}_{\rm Cep}$}  &    
 \colhead{$\mu^{0}_{\rm TRGB}$} &    
 \colhead{Source}               & 
 \colhead{$\Delta$}             &
 \colhead{$\mu^{0}_{\rm TRGB}$} &    
 \colhead{$\Delta$}             &   
 \colhead{Source}   
\\
 & & \colhead{$M_{I}=-4.05$} & & & \colhead{(met. corr.)} & &
\\
 \colhead{(1)}     & 
 \colhead{(2)}     & 
 \colhead{(3)}     & 
 \colhead{(4)}     & 
 \colhead{(5)}     & 
 \colhead{(6)}     & 
 \colhead{(7)}     & 
 \colhead{(8)}     
} 
\startdata
NGC\,55   & 26.41 & 26.64 & a,b & $-0.23$ & \nodata & \nodata   &     \\
NGC\,224  & 24.54 & 24.46 & c,d & $+0.08$ & 24.37   & $+0.17$   & 1   \\
NGC\,300  & 26.48 & 26.56 & d   & $-0.08$ & 26.48   & $+0.00$   & 1   \\
NGC\,598  & 24.64 & 24.66 & c,d & $-0.02$ & 24.71   & $-0.07$   & 1   \\
NGC\,2403 & 27.43 &(27.62)& e   &($-0.19$)& \nodata & \nodata   &     \\   
NGC\,3031 & 27.80 & 27.80 & d   & $+0.00$ & 27.70   & $+0.10$   & 1   \\
NGC\,3109 & 25.45 & 25.54 & d   & $-0.09$ & 25.57   & $-0.12$   & 1   \\
NGC\,3351 & 30.10 & 30.23 & d,f & $-0.13$ & 29.92   & $+0.18$   & 1   \\
NGC\,3621 & 29.30 & 29.27 & d   & $+0.03$ & 29.26   & $+0.04$   & 1   \\
NGC\,4258 & 29.50 & 29.32 & g,h & $+0.18$ & 29.37   & $+0.13$   & 2,3 \\
NGC\,5128 & 27.67 & 27.89 & i,j & $-0.22$ & 27.90   & $-0.23$   & 4   \\
NGC\,5236 & 28.32 & 28.56 & k   & $-0.24$ & \nodata & \nodata   &     \\
NGC\,5253 & 28.05 & 27.89 & f   & $+0.16$ & \nodata & \nodata   &     \\ 
NGC\,5457 & 29.17 & 29.39 & d,f & $-0.22$ & 29.34   & $-0.17$   & 1   \\
NGC\,6822 & 23.31 & 23.37 & f   & $-0.09$ & 23.37   & $-0.06$   & 5   \\
IC\,1613  & 24.32 & 24.33 & d   & $-0.01$ & 24.38   & $-0.06$   & 1   \\
IC\,4182  & 28.21 & 28.19 & d   & $+0.02$ & 28.23   & $-0.02$   & 1   \\
WLM       & 24.82 & 24.87 & c,d & $-0.05$ & 24.93   & $-0.11$   & 1   \\ 
\tableline
{\bf mean} &       &       & & 
\multicolumn{2}{l}{\boldmath{$-0.05\pm0.03$}} &
\multicolumn{2}{l}{\boldmath{$-0.02\pm0.03$}} \\
 &       &       & & 
\multicolumn{2}{l}{$\sigma=0.13,\,N=17$} &
\multicolumn{2}{l}{$\sigma=0.13,\,N=14$} 
\enddata
\tablenotetext{1)}{Added are here NGC\,55, IC\,1613, and WLM from \S~\ref{sec:03:3}.}
\tablenotetext{\,}{References to $\mu^{0}_{\rm TRGB}$:
   (a) \citealt{Seth:etal:05};
   (b) \citealt{Tully:etal:06};
   (c) \citealt{McConnachie:etal:05};
   (d) \citealt{Rizzi:etal:07};
   (e) mean distance of 6 group members;
   (f) \citealt{Sakai:etal:04};
   (g) \citealt{Mouhcine:etal:05};
   (h) \citealt{Macri:etal:06};
   (i) \citealt{Rejkuba:etal:05};
   (j) \citealt{Karataeva:etal:06}
   (k) \citealt{Karachentsev:etal:07}
}
\tablenotetext{\,}{References to metallicity-corrected $\mu^{0}_{\rm
    TRGB}$: 
   (1) \citealt{Rizzi:etal:07};
   (2) \citealt{Macri:etal:06};
   (3) \citealt{Mouhcine:etal:05};
   (4) \citealt{Ferrarese:etal:07};
   (5) \citealt{Sakai:etal:04}. 
   Some additional distances:} 
\tablenotetext{\,}{
NGC\,224 ~~  $\mu^{0}=24.44\pm0.12$ from an eclipsing binary
    \citep{Ribas:etal:05}}
\tablenotetext{\,}{
NGC\,300 ~~  $\mu^{0}=26.41$ from $VIJK$ photometry of Cepheids and
    assuming $\mu^{0}_{\rm LMC}=18.54$ \citep{Gieren:etal:05a}}
\tablenotetext{\,}{
NGC\,598 ~~ $\mu^{0}=24.92\pm0.12$ from an eclipsing binary
    \citep{Bonanos:etal:06}} 
\tablenotetext{\,}{
NGC\,4258 ~  $\mu^{0}=29.29\pm0.08\pm0.07$ from water maser
    \citep{Herrnstein:etal:99}} 
\tablenotetext{\,}{
NGC\,6822 ~ $\mu^{0}=23.35$ from $VIJK$ photometry of Cepheids and
    assuming $\mu^{0}_{\rm LMC}=18.54$
    \citep{Gieren:etal:06}} 
\end{deluxetable}
\setlength\textheight{8.4in}%

\clearpage


\begin{thebibliography}{}
%
\bibitem[Alcock et~al.(2004)]{Alcock:etal:04}
   Alcock, C., et~al. 2004, 
   AJ, 127, 334   (LMC)
%
\bibitem[Alves(2004)]{Alves:04}
   Alves, D.~R. 2004,
   New Astron. Rev., 48, 659
%
\bibitem[An et~al.(2007)]{An:etal:07}
   An, D., Terndrup, D.~M., \& Pinsonneault, M.~H. 2007,
   ApJ, in press (astro-ph/0707.3144) 
%
\bibitem[Antonello et~al.(2006)]{Antonello:etal:06}
   Antonello, E., Fossati, L., Fugazza, D., Mantegazza, L., \& Gieren,
   W. 2006,
   A\&A, 445, 901 (IC\,1613)
%
\bibitem[Aparicio et~al.(2001)]{Aparicio:etal:01}
   Aparicio, A., Carrera, R., \& Mart{\'i}nez-Delgado, D. 2001, 
   AJ, 122, 2524   (Draco)
%
\bibitem[Baade(1944a)]{Baade:44a}
   Baade, W. 1944a,
   ApJ, 100, 137
%
\bibitem[Baade(1944b)]{Baade:44b}
   Baade, W. 1944b,
   ApJ, 100, 147
%
\bibitem[Baade(1948)]{Baade:48}
   Baade, W. 1948,
   PASP, 60, 230
%
\bibitem[Baade(1954)]{Baade:54}
   Baade, W. 1954,
   %
   in Trans. IAU, VIII (Rome 1952 meeting), 
   Report of Commission 28, 
   Cambridge Univ. Press.
%
\bibitem[Barnes et~al.(2003)]{Barnes:etal:03}
   Barnes, T., Jeffreys, W., Berger, J., Mueller, P., Orr, K., \&
   Rodriguez, R. 2003,
   ApJ, 592, 539
%
\bibitem[Bauer et~al.(1999)]{EROS:99}
   Bauer, F., et~al. (EROS Collaboration) 1999, 
   A\&A, 348, 175
%
\bibitem[Behr(1951)]{Behr:51}
   Behr, A. 1951, 
   AN, 279, 97
%
\bibitem[Bellazzini et~al.(2002)]{Bellazzini:etal:02}
   Bellazzini, M., Ferraro, F.~R., Origlia, L., Pancino, E., 
   Monaco, L., \& Oliva, E. 2002, 
   AJ, 124, 3222 (UMi)
%
\bibitem[Bellazzini et~al.(2001)]{Bellazzini:etal:01} 
   Bellazzini, M., Ferraro, F.~R., \& Pancino, E. 2001, 
   ApJ, 556, 635
%
\bibitem[Bellazzini et~al.(2004a)]{Bellazzini:etal:04a}
   Bellazzini, M., Ferraro, F.~R., Sollima, A., Pancino, E., \&
   Origlia, L. 2004a, 
   A\&A, 424, 199
%
\bibitem[Bellazzini et~al.(2004b)]{Bellazzini:etal:04b}
   Bellazzini, M., Gennari, N., Ferraro, F.~R., Sollima, A. 2004b, 
   MNRAS, 354, 708
%
\bibitem[Bencivenni et~al.(1991)]{Bencivenni:etal:91}
   Bencivenni, D., Caputo, F., Manteign, M., \& Quarta, M.~L. 1991,
   ApJ, 380, 484
%
\bibitem[Benedict et~al.(2002)]{Benedict:etal:02} 
   Benedict, G.~F., et~al. 2002,
   AJ, 124, 1695
%
\bibitem[Benedict et~al.(2007)]{Benedict:etal:07} 
   Benedict, G.~F., et~al. 2007,
   AJ, 133, 1810 
%
\bibitem[Berdnikov et~al.(2000)]{Berdnikov:etal:00}
   Berdnikov, L.~N., Dambis, A.~K., \& Voziakova, O.~V. 2000,
   A\&AS, 143, 211
%
\bibitem[Bergbusch \& VandenBerg(2001)]{Bergbusch:VandenBerg:01}
   Bergbusch, P.~A., \& VandenBerg, D.~A. 2001,
   ApJ, 556, 322
%
\bibitem[Bersier(2000)]{Bersier:00}
   Bersier, D. 2000,
   ApJ, 543, L23
%
\bibitem[Bersier \& Wood(2002)]{Bersier:Wood:02}
   Bersier, D., \& Wood, P.~R. 2002,
   AJ, 123, 840   (Fornax)
%
\bibitem[Bonanos et~al.(2004)]{Bonanos:etal:04}
   Bonanos, A.~Z., Stanek, K.~Z., Szentgyorgyi, A.~H., Sasselov, D.~D., \&
   Bakos, G.~A. 2004, 
   AJ, 127, 861   (Draco)
%
\bibitem[Bonanos et~al.(2006)]{Bonanos:etal:06}
   Bonanos, A.~Z., et~al. 2006, 
   ApJ, 652, 313
%
\bibitem[Bono et~al.(2007)]{Bono:etal:07}
   Bono, G., Caputo, F., \& di~Criscienzo, M. 2007, 
   A\&A, 476, 779 
%
\bibitem[Borissova et~al.(2004)]{Borissova:etal:04}
   Borissova, J., Minniti, D., Rejkuba, M., Alves, D., Cook, K.~H., \&
   Freeman, K.~C. 2004, 
   A\&A, 423, 97   (LMC)
%
\bibitem[Brown et~al.(2004)]{Brown:etal:04}
   Brown, T.~M., Ferguson, H.~C., Smith, E., Kimble, R.~A., Sweigart,
   A.~V., Renzini, A., \& Rich, R.~M. 2004, 
   AJ, 127, 2738   (NGC\,224)
%
\bibitem[Caldwell(2006)]{Caldwell:06}
   Caldwell, N. 2006, 
   ApJ, 651, 822
%
\bibitem[Caloi et~al.(1997)]{Caloi:etal:97}
   Caloi, V., D'Antona, F., \& Mazzitelli, I. 1997, 
   A\&A, 320, 823
%
\bibitem[Caputo(1997)]{Caputo:97}
   Caputo, F. 1997, 
   MNRAS, 284, 994
%
\bibitem[Caputo et~al.(2000)]{Caputo:etal:00}
   Caputo, F., Castellani, V., Marconi, M., \& Ripepi, V. 2000,
   MNRAS, 316, 819
%
\bibitem[Caputo et~al.(1993)]{Caputo:etal:93}
   Caputo, F., de Rinaldis, A., Manteiga, M., 
   Pulone, L., \& Quarta, M.~L. 1993,
   A\&A, 276, 41
%
\bibitem[Carrera et~al.(2002)]{Carrera:etal:02}
   Carrera, R., Aparicio, A., Mart{\'i}nez-Delgado, D., \&
   Alonso-Garc{\'i}a, J. 2002,
   AJ, 123, 3199   (UMi)
%
\bibitem[Carretta \& Gratton(1997)]{Carretta:Gratton:97}
   Carretta, E., \& Gratton, R.~G. 2000,
   A\&AS, 121, 95
%
\bibitem[Carretta et~al.(2000)]{Carretta:etal:00}
   Carretta, E., Gratton, R.~G., Clementini, G., \& Fusi Peci, F. 2000,
   ApJ, 533, 215
%
\bibitem[Cassisi et~al.(1999)]{Cassisi:etal:99}
   Cassisi, S., Castellani, V., degl'Innocenti, S., 
   Salaris, M., \& Weiss, A. 1999,
   A\&AS, 134, 103
%
\bibitem[Castellani et~al.(1991)]{Castellani:etal:91}
   Castellani, V., Chieffi, S., \& Pulone, L. 1991,
   ApJS, 76, 911
%
\bibitem[Catelan et~al.(2004)Catelan, Pritzl, \& Smith]{Catelan:etal:04}
   Catelan, M., Pritzl, B.~J., \& Smith, H.~A. 2004, 
   ApJS, 154, 633
%
\bibitem[Cioni et~al.(2000)]{Cioni:etal:00}
   Cioni, M.-R.~L., van der Marel, R.~P., Loup, C., \& Habing, H.~J. 2000,
   A\&A, 359, 601
%
\bibitem[Clausen et~al.(2003)]{Clausen:etal:03}
   Clausen, J.~V., Storm, J., Larsen, S.~S., \& Gim{\'e}nez, A. 2003,
   A\&A, 402, 509
%
\bibitem[Clementini et~al.(2003a)]{Clementini:etal:03a}
   Clementini, G., Gratton, R., Bragaglia, A., Carretta, E., Di
   Fabrizio, L., \& Maio, M. 2003a,
   AJ, 125, 1309   (LMC)
%
\bibitem[Clementini et~al.(2003b)]{Clementini:etal:03b}
   Clementini, G., Held, E.~V., Baldacci, L., \& Rizzi, L. 2003b,
   ApJ, 588, L85   (NGC\,6822)
%
\bibitem[Da~Costa \& Armandroff(1990)]{DaCosta:Armandroff:90}
   Da Costa, G.~S., \& Armandroff, T.~E. 1990,
   AJ, 100, 162
%
\bibitem[Dale \& Giovanelli(2000)]{Dale:Giovanelli:00}
   Dale, D.~A., \& Giovanelli, R. 2000,
   in Cosmic Flows Workshop, eds. S. Courteau \& J. Willick, ASP
   Conf. Ser. 201, 25
%
\bibitem[Dall'Ora et~al.(2004)]{Dall'Ora:etal:04}
   Dall'Ora, M., et~al. 2004, 
   ApJ, 610, 269
%
\bibitem[de Freitas Pacheco(1986)]{deFreitasPacheco:86}
   de Freitas Pacheco, J.~A. 1986, 
   Rev. Mex. Astron. Astrofis., 12, 74 
%
\bibitem[Demarque et~al.(2000)]{Demarque:etal:00}
   Demarque, P., Zinn, R., Lee, Y.~W., \& Yi, S. 2000, 
   AJ, 119, 1398
%
\bibitem[Demers \& Irwin(1993)]{Demers:Irwin:93}
   Demers, S., \& Irwin, M.~J. 1993,
   MNRAS, 261, 657   (Leo~II)
%
\bibitem[De~Santis \& Cassisi(1999)]{DeSantis:Cassisi:99}
   De~Santis, R., \& Cassisi, S. 1999, 
   MNRAS, 308, 97
%
\bibitem[Dolphin(2000)]{Dolphin:00}
   Dolphin, A.~E. 2000,
   ApJ, 531, 804
%
\bibitem[Dolphin et~al.(2001)]{Dolphin:etal:01}
   Dolphin, A.~E., et~al. 2001, 
   ApJ, 550, 554   (IC\,1613)
%
\bibitem[Dolphin et~al.(2002)]{Dolphin:etal:02}
   Dolphin, A.~E., et~al. 2002, 
   AJ, 123, 3154   (Leo~A)
%
\bibitem[Dolphin et~al.(2003)]{Dolphin:etal:03}
   Dolphin, A.~E., et~al. 2003, 
   AJ, 125, 1261
%
\bibitem[Dorman(1992)]{Dorman:92}
   Dorman, B. 1992, 
   ApJS, 81, 221
%
\bibitem[Dressler(1984)]{Dressler:84}
   Dressler, A. 1984, 
   ApJ, 281, 512
%
\bibitem[Durrell et~al.(2002)]{Durrell:etal:02}
   Durrell, P.~R., Ciardullo, R., Feldmeier, J.~J., Jacoby, G.~H., \&
   Sigurdsson, S. 2002, 
   ApJ, 570, 119
%
\bibitem[Durrell et~al.(2007)]{Durrell:etal:07}
   Durrell, P.~R., et~al. 2007, 
   ApJ, 656, 746 
%
\bibitem[Feast(1999)]{Feast:99}
   Feast, M.~W. 1999,
   PASP, 111, 775 
%
\bibitem[Feast(2004)]{Feast:04}
   Feast, M.~W. 2004,
   in Variable Stars in the Local Group,
   eds. D.~W. Kurtz \& K.~R. Pollard
   (San Francisco: ASP), 304
%
\bibitem[Federspiel et~al.(1994)]{Federspiel:etal:94}
   Federspiel, M., Sandage, A., \& Tammann, G.~A. 1994,
   ApJ, 430, 29
%
\bibitem[Fernie(1990)]{Fernie:90}
   Fernie, J.~D. 1990,
   ApJS, 72, 153
%
\bibitem[Fernie et~al.(1995)]{Fernie:etal:95}
   Fernie, J.~D., Beattie, B., Evans, N.~R., \& Seager, S. 1995,
   IBVS, 4148  (http://ddo.astro.utoronto.ca/cepheids.html)
%
\bibitem[Ferrarese et~al.(2007)]{Ferrarese:etal:07}
   Ferrarese, L., Mould, J.~R., Stetson, P.~B., Tonry, J.~L., Blakeslee,
   J.~P., \&  Ajhar, E.~A. 2007,
   ApJ, 654, 186
%
\bibitem[Ferrarese et~al.(1996)]{Ferrarese:etal:96}
   Ferrarese, L., et~al. 1996, 
   ApJ, 464, 568 (NGC\,4321)
%
\bibitem[Ferrarese et~al.(2000a)]{Ferrarese:etal:00}
   Ferrarese, L., et~al. 2000a, 
   ApJS, 128, 431
%
\bibitem[Ferrarese et~al.(2000b)]{Ferrarese:etal:00b}
   Ferrarese, L., et~al. 2000b, 
   ApJ, 529, 745
%
\bibitem[Ferraro et~al.(1999)]{Ferraro:etal:99}
   Ferraro, F.~R., Messineo, M., Fusi Pecci, F., de~Palo, M.~A.,
   Stranieero, O., Chieffi, A., \& Limongi, M. 1999,
   AJ, 118, 1738
%
\bibitem[Fitzpatrick et~al.(2002)]{Fitzpatrick:etal:02}
   Fitzpatrick, E.~L., Ribas, I., Guinan, E.~F., De Warf, L.~E., Maloney,
   F.~P., \& Massa, D. 2002,
   ApJ, 564, 260
%
\bibitem[Fouqu{\'e} et~al.(2003)]{Fouque:etal:03}
   Fouqu{\'e}, P., Storm, J., \& Gieren, W. 2003,
   Lect. Notes Phys., 635, 21
%
\bibitem[Fouqu{\'e} et~al.(2007)]{Fouque:etal:07}
   Fouqu{\'e}, P., et~al. 2007,
   A\&A, 476, 73 
%
\bibitem[Freedman et~al.(2001)]{Freedman:etal:01}
   Freedman, W.~L., et~al. 2001, 
   ApJ, 553, 47
%
\bibitem[Fusi~Pecci et~al.(1996)]{Fusi-Pecci:etal:96}
   Fusi~Pecci, F., et~al. 1996, 
   AJ, 112, 1461
%
\bibitem[Gallart et~al.(1996)]{Gallart:etal:96}
   Gallart, C., Aparicio, A., Vilchez, J.~M. 1996, 
   AJ, 112, 1928
%
\bibitem[Gallart et~al.(2004)]{Gallart:etal:04}
   Gallart, C., Aparicio, A., Freedman, W.~L., Madore, B.~F.,
   Mart{\'i}nez-Delgado, D., \& Stetson, P.~B. 2004, 
   AJ, 127, 1486   (Phoenix)
%
\bibitem[Galleti et~al.(2004)]{Galleti:etal:04}
   Galleti, S., Bellazzini, M., \& Ferraro, F.~R. 2004, 
   A\&A, 423, 925
%
\bibitem[Gascoigne \& Kron(1965)]{Gascoigne:Kron:65}
   Gascoigne, S.~C.~B., \& Kron, G.~E. 1965, 
   MNRAS, 130, 933
%
\bibitem[Gibson et~al.(2000)]{Gibson:etal:00}
   Gibson, B.~K., et~al.  2000, 
   ApJ, 529, 723
%
\bibitem[Gieren et~al.(2005a)]{Gieren:etal:05a}
   Gieren, W., Pietrzynski, G., Soszynski, I., Bresolin, F.,
    Kudritzki, R.-P., Minniti, D., \& Storm, J. 2005a,
   ApJ, 628, 695
%
\bibitem[Gieren et~al.(2005b)]{Gieren:etal:05b}
   Gieren, W., Storm, J., Barnes, T.~G., Fouqu{\'e}, P., Pietrzynski,
   G., \& Kienzle, F. 2005b,
   ApJ, 627, 224
%
\bibitem[Gieren et~al.(2006)]{Gieren:etal:06}
   Gieren, W., Pietrzynski, G., Nalewajko, K., Soszynski, I.,
   Bresolin, F., Kudritzki, R.-P., Minniti, D., \& Romanowsky, A. 2006,
   ApJ, 647, 1056
%
\bibitem[Giraud(1990)]{Giraud:90}
   Giraud, E. 1990,
   A\&A, 231, 1
%
\bibitem[Graham et~al.(1997)]{Graham:etal:97}
   Graham, J.~A., et~al. 1997, 
   ApJ, 477, 535 (NGC\,3351)
%
\bibitem[Gratton et~al.(1997)]{Gratton:etal:97}
   Gratton, R.~G., Fusi Pecci, F., Carretta, E., Clementini, G.,
   Corsi, C.~E., \& Lattanzi, M. 1997, 
   ApJ, 491, 749
%
\bibitem[Gratton(1998)]{Gratton:98}
   Gratton, R.~G. 1998, 
   MNRAS, 296, 739
%
\bibitem[Greco et~al.(2005)]{Greco:etal:05}
   Greco, C., et~al. 2005,
   in Resolved Stellar Populations, 
   eds. M.~Chavez \& D.~Valls-Gabaud
   (San Francisco: ASP), (astro-ph/0507244)  (Fornax)
%
\bibitem[Grillmair et~al.(1998)]{Grillmair:etal:98}
   Grillmair, C.~J., et~al. 1998, 
   AJ, 115, 144   (Draco)
%
\bibitem[Groenewegen \& Salaris(2003)]{Groenewegen:Salaris:03}
   Groenewegen, M.~A.~T., \& Salaris, M. 2003,
   A\&A, 410, 887
%
\bibitem[Han et~al.(1997)]{Han:etal:97}
   Han, H., Hoessel, J.~G., Gallagher, J.~S., Holtsman, J., \& 
   Stetson, P.~B. 1997, 
   AJ, 113, 1001
%
\bibitem[Held et~al.(2001)]{Held:etal:01}
   Held, E.~V., Clementini, G., Rizzi, L., Momany, Y., Saviane, I., \&
   Di Fabrizio, L. 2001,
   ApJ, 562, L39   (Leo~I)
%
\bibitem[Held et~al.(2000)]{Held:etal:00}
   Held, E.~V., Saviane, I., Momany, Y., \& Carraro, G. 2000,
   ApJ, 530, L85 (Leo~I)
%
\bibitem[Herrnstein et~al.(2005)]{Herrnstein:etal:05}
   Herrnstein, J.~R., Moran, J.~M., Greenhill, L.~J., \& Trotter, A.~S. 2005, 
   ApJ, 629, 719
%
\bibitem[Herrnstein et~al.(1999)]{Herrnstein:etal:99}
   Herrnstein, J.~R., et~al. 1999, 
   Nature, 400, 539
%
\bibitem[Hertzsprung(1913)]{Hertzsprung:13}
   Hertzsprung, E. 1913,
   AN, 196, 201
%
\bibitem[Hilditch et~al.(2005)]{Hilditch:etal:05}
   Hilditch, R.~W., Howarth, I.~D., \& Harries, T.~J. 2005,
   MNRAS, 357, 304
%
\bibitem[Holmberg(1950)]{Holmberg:50}
   Holmberg, E. 1950, 
   Medd. Lunds Obs. 128, 1
%
\bibitem[Hough et~al.(1987)]{Hough:etal:87}
   Hough, J.~H., Bailey, J.~A., Rouse, M.~F., \& Whittet, D.~C.~B. 1987, 
   MNRAS, 227, 1
%
\bibitem[Hubble(1925)]{Hubble:25}
   Hubble, E. 1925, 
   ApJ, 62, 409 (NGC\,6822)
%
\bibitem[Hubble(1926)]{Hubble:26}
   Hubble, E. 1926, 
   ApJ, 63, 236 (M\,33)
%
\bibitem[Hubble(1929)]{Hubble:29}
   Hubble, E. 1929, 
   ApJ, 69, 103 (M\,31)
%
\bibitem[Hubble(1951)]{Hubble:51}
   Hubble, E. 1951, 
   %
   Trans. Am. Phil. Soc. 51, 461 (The Penrose Lecture)
%
\bibitem[Hubble \& Sandage(1953)]{Hubble:Sandage:53}
   Hubble, E., \& Sandage, A. 1953, 
   ApJ, 118, 353
%
\bibitem[Jerjen \& Tammann(1993)]{Jerjen:Tammann:93}
   Jerjen, H., \& Tammann, G.~A. 1993,
   A\&A, 276, 1
%
\bibitem[Kaluzny et~al.(1995)]{Kaluzny:etal:95}
   Kaluzny, J., Kubiak, M., Szymanski, M., Udalski, A., Krzeminski,
   W., \& Mateo, M. 1995, 
   ApJS, 112, 407   (Sculptor)
%
\bibitem[Kanbur et~al.(2007)]{Kanbur:etal:07}
   Kanbur, S.~M., Ngeow, C., Nanthakumar, A., \& Stevens, R. 2007, 
   PASP, 119, 512 
%
\bibitem[Karachentsev(2005)]{Karachentsev:05}
   Karachentsev, I.~D. 2005, 
   AJ, 129, 178
%
\bibitem[Karachentsev et~al.(2004)]{Karachentsev:etal:04}
   Karachentsev, I.~D., Karachentseva, V.~E., Huchtmeier, W.~K., 
   \& Makarov, D.~I. 2004, 
   AJ, 127, 2031
%
\bibitem[Karachentsev et~al.(2003)]{Karachentsev:etal:03}
   Karachentsev, I.~D., et~al. 2003, 
   A\&A, 398, 467
%
\bibitem[Karachentsev et~al.(2006)]{Karachentsev:etal:06} 
   Karachentsev, I.~D., et~al. 2006, 
   AJ, 131, 1361 
%
\bibitem[Karachentsev et~al.(2007)]{Karachentsev:etal:07} 
   Karachentsev, I.~D., et~al. 2007, 
   AJ, 133, 504 
%
\bibitem[Karataeva et~al.(2006)]{Karataeva:etal:06} 
   Karataeva, G.~M., Tikhonov, N.~A., Galazutdinova, O.~A., 
   \& Hagen-Thorn, V.~A. 2006, 
   Astron. Letters 32, 236
%
\bibitem[Keller \& Wood(2006)]{Keller:Wood:06}
   Keller, S.~C., \& Wood, P.~R. 2006,
   ApJ, 642, 834
%
\bibitem[Kennicutt et~al.(2003)]{Kennicutt:etal:03}
   Kennicutt, R.~C., Bresolin, F., \& Garnett, D.~R. 2003, 
   ApJ, 591, 801
%
\bibitem[Kervella et~al.(2004)]{Kervella:etal:04}
   Kervella, P., Nardetto, N., Bersier, D., Mourard, D., \& Coud{\'e}
   du Foresto, V. 2004,
   A\&A, 416, 941
%
\bibitem[Klypin et~al.(2003)]{Klypin:etal:03}
   Klypin, A., Hoffman, Y., Kravtsov, A.~V., \& Gottl{\"o}ber, S. 2003,
   ApJ, 596, 19
%
\bibitem[Koen et~al.(2007)]{Koen:etal:07}
   Koen, C., Kanbur, S., \& Ngeow, C. 2007,
   MNRAS, 380, 1440
%
\bibitem[Koen \& Siluyele(2007)]{Koen:Siluyele:07}
   Koen, C., \& Siluyele, I. 2007,
   MNRAS, 377, 1281
%
\bibitem[Kraan-Korteweg(1986)]{Kraan-Korteweg:86}
   Kraan-Korteweg, R.~C. 1986,
   A\&AS, 66, 255
%
\bibitem[Lane et~al.(2002)]{Lane:etal:02}
   Lane, B., Creech-Eakman, M., \& Nordgren, T. 2002,
   ApJ, 573, 330
%
\bibitem[Laney \& Stobie(1986)]{Laney:Stobie:86}
   Laney, C.~D., \& Stobie, R.~S. 1986,
   MNRAS, 222, 449
%
\bibitem[Layden \& Sarajedini(2000)]{Layden:Sarajedini:00}
   Layden, A.~C., \& Sarajedini, A. 2000, 
   AJ, 119, 1760   (Sag dSph)
%
\bibitem[Lee et~al.(1993)]{Lee:etal:93}
   Lee, M.~G., Freedman, W.~L., \& Madore, B.~F. 1993, 
   ApJ, 417, 553 
%
\bibitem[Lee et~al.(2003)]{Lee:etal:03}
   Lee, M.~G., et~al. 2003, 
   AJ, 126, 2840
%
\bibitem[Lee et~al.(1990)]{Lee:etal:90}
   Lee, Y.~W., Demarque, P., \& Zinn, R. 1990, 
   ApJ, 350, 155
%
\bibitem[Macri et~al.(2006)]{Macri:etal:06}
   Macri, L.~M., Stanek, K.~Z., Bersier, D., Greenhill, L.~J., \& 
   Reid, M.~J. 2006,
   ApJ, 652, 1133
%
\bibitem[Mackey \& Gilmore(2003)]{Mackey:Gilmore:03}
   Mackey, A.~D., \& Gilmore, G.~F. 2003,
   MNRAS, 345, 747   (Fornax)
%
\bibitem[Madore \& Freedman(1991)]{Madore:Freedman:91}
   Madore, B.~F., \& Freedman, W.~L. 1991,
   PASP, 103, 933
%
\bibitem[Marconi et~al.(2005)]{Marconi:etal:05}
   Marconi, M., Musella, I., \& Fiorentino, G. 2005,
   ApJ, 632, 590
%
\bibitem[Masters et~al.(2006)]{Masters:etal:06}
   Masters, K.~L., Springob, C.~M., Haynes, M.~P., \& Giovanelli, R. 2006,
   ApJ, 653, 861
%
\bibitem[Mateo et~al.(1995a)]{Mateo:etal:95a}
   Mateo, M., Fischer, P., \& Krzeminski, W. 1995a,
   AJ, 110, 2166   (Sextans)
%
\bibitem[Mateo et~al.(1995b)]{Mateo:etal:95b}
   Mateo, M., Kubiak, M., Szymanski, M., Kaluzny, J., Krzeminski, W.,
   \& Udalski, A. 1995b,
   AJ, 110, 1141   (Sag dSph)
%
\bibitem[McAlary et~al.(1983)]{McAlary:etal:83}
   McAlary, C.~W., Madore, B.~F., McGonegal, R., McLaren, R.~A., \&
   Welch, D.~L. 1983,
   ApJ, 273, 539   (NGC\,6822)
%
\bibitem[McConnachie et~al.(2004)]{McConnachie:etal:04}
   McConnachie, A.~W., Irwin, M.~J., Ferguson, A.~M.~N., Ibata, R.~A.,
   Lewis, G.~F., \& Tanvir, N. 2004, 
   MNRAS, 350, 243
%
\bibitem[McConnachie et~al.(2005)]{McConnachie:etal:05}
   McConnachie, A.~W., Irwin, M.~J., Ferguson, A.~M.~N., Ibata, R.~A.,
   Lewis, G.~F., \& Tanvir, N. 2005, 
   MNRAS, 356, 979
%
\bibitem[McNamara(1997)]{McNamara:97}
   McNamara, D.~H. 1997, 
   PASP, 109, 1221
%
\bibitem[McNamara et~al.(2004)]{McNamara:etal:04}
   McNamara, D.~H., Rose, M.~B., Brown, P.~J., Ketcheson, D.~I., 
   Maxwell, J.~E., Smith, K.~M., \&  Wooley, R.~C. 2004,
   in Variable Stars in the Local Group,
   eds. D.~W. Kurtz \& K.~R. Pollard
   (San Francisco: ASP), 525
%
\bibitem[Mei et~al.(2007)]{Mei:etal:07}
   Mei, S., et~al. 2007,
   ApJ, 655, 144
%
\bibitem[Mouhcine et~al.(2005)]{Mouhcine:etal:05}
   Mouhcine, M., Ferguson, H.~C., Rich, R.~M., Brown, T.~M., \& 
   Smith, T.~E. 2005,
   ApJ, 633, 810
%
\bibitem[Nardetto et~al.(2004)]{Nardetto:etal:04}
   Nardetto, N., Fokin, A., Mourard, D., Mathias, P., Kervella, P.,
   \& Bersier, D. 2004, 
   A\&A, 428, 131
%
\bibitem[Natale et~al.(2007)]{Natale:etal:07}
   Natale, G., Marconi, M., \& Bono, G. 2007,
   preprint, astro-ph/0711.2857
%
\bibitem[Nemec(1985)]{Nemec:85}
   Nemec, J.~M. 1985,
   AJ, 90, 204   (Draco)
%
\bibitem[Nemec(1989)]{Nemec:89}
   Nemec, J.~M. 1989, 
   in The Use of Variable Stars in Fundamental Problems of Astronomy, 
   ed. E.~P. Schmidt
   (Cambridge: Cambridge Univ. Press), 215
%
\bibitem[Nemec \& Mateo(1990a)]{Nemec:Mateo:90a}
   Nemec, J.~M., \& Mateo, M. 1990a,
   in The Evolution of the Universe of Galaxies, 
   ed. R.~G. Kron
   (San Francisco: ASP), 134
%
\bibitem[Nemec \& Mateo(1990b)]{Nemec:Mateo:90b}
   Nemec, J.~M., \& Mateo, M. 1990b,
   in Confrontation between stellar pulsation and evolution, 
   (San Francisco: ASP), 64
%
\bibitem[Nemec et~al.(1988)]{Nemec:etal:88}
   Nemec, J.~M., Wehlau, A., \& Mendes de Oliveira, C. 1988,
   AJ, 96, 528   (UMi)
%
\bibitem[Ngeow et~al.(2005)]{Ngeow:etal:05} 
   Ngeow, C.-C., Kanbur, S.~M., Nokolaev, S., Buonacorsi, J., Cook, K.~H., 
   \& Welch, D.~L. 2005,
   MNRAS, 363, 831
%
\bibitem[Nordgren et~al.(2002)]{Nordgren:etal:02}
   Nordgren, T.~E., Lane, B.~F., Hindsley, R.~B., \& Kervella, P. 2002,
   AJ, 123, 3380
%
\bibitem[Panagia(2005)]{Panagia:05}
   Panagia, N. 2005,
   in Cosmic Explosions,
   eds. J.~M. Marcaide \& K.~W. Weiler
   (Berlin: Springer), 585
%
\bibitem[Pietrinferni et~al.(2004)]{Pietrinferni:etal:04}
   Pietrinferni, A., Cassisi, S., Salaris, M., \& Castelli, F. 2004, 
   ApJ, 612, 168
%
\bibitem[Pietrzynski et~al.(2004)]{Pietrzynski:etal:04}
   Pietrzynski, G., Gieren, W., Udalski, A., Bresolin, F., 
   Kudritzki, R.-P., Soszynski, I., Szymanski, M., \&  Kubiak, M. 2004, 
   AJ, 128, 2815 (NGC\,6822)
%
\bibitem[Pietrzynski et~al.(2006a)]{Pietrzynski:etal:06a}
   Pietrzynski, G., et~al. 2006a, 
   AJ, 132, 2556 (NGC\,55)
%
\bibitem[Pietrzynski et~al.(2006b)]{Pietrzynski:etal:06b}
   Pietrzynski, G., et~al. 2006b, 
   ApJ, 648, 366 (NGC\,3109)
%
\bibitem[Pietrzynski et~al.(2007)Pietrzynski's et~al.]{Pietrzynski:etal:07}
   Pietrzynski, G., et~al. 2007, 
   AJ, 134, 594 (WLM) 
%
\bibitem[Piotto et~al.(1994)]{Piotto:etal:94}
   Piotto, G., Capaccioli, M., \& Pellegrini, C. 1994,
   A\&A, 287, 371 (Sextans A+B)
%
\bibitem[Popowski \& Gould(1999)]{Popowski:Gould:99}
   Popowski, P., \& Gould, A. 1999,
   in Post Hipparcos Standard Candles,
   eds. A.~Heck, \& F.~Caputo
   (Dordrecht: Kluwer), 53
%
\bibitem[Pritzl et~al.(2002)]{Pritzl:etal:02}
   Pritzl, B.~J., Armandroff, T.~E., Jacoby, G.~H., \& Da Costa, G.~S. 2002,
   AJ, 124, 1464   (And~VI)
%
\bibitem[Pritzl et~al.(2004)]{Pritzl:etal:04}
   Pritzl, B.~J., Armandroff, T.~E., Jacoby, G.~H., \& Da~Costa, G.~S. 2004, 
   AJ, 127, 318   (And~II)
%
\bibitem[Pritzl et~al.(2005)]{Pritzl:etal:05}
   Pritzl, B.~J., Armandroff, T.~E., Jacoby, G.~H., \& Da Costa, G.~S. 2005,
   AJ, 129, 2232   (And (I\&III)
%
\bibitem[Reid(1997)]{Reid:97}
   Reid, I.~N. 1997, 
   AJ, 114, 161
%
\bibitem[Reid(1999)]{Reid:99}
   Reid, I.~N. 1999, 
   ARA\&A, 37, 191
%
\bibitem[RTS\,05(2005)Reindl et~al.]{RTS:05}
   Reindl, B., Tammann, G.~A., Sandage, A., \& Saha, A. 2005,
   ApJ, 624, 532 (RTS\,05)
%
\bibitem[Rejkuba et~al.(2005)]{Rejkuba:etal:05}
   Rejkuba, M., Greggio, L., Harris, W.~E., Harris, G.~L.~H., \&
   Peng, E.~W. 2005,
   ApJ, 631, 262
%
\bibitem[Rejkuba et~al.(2000)]{Rejkuba:etal:00}
   Rejkuba, M., Minniti, D., Gregg, M.~D., Zijlstra, A.~A., Alonso,
   M.~V., \& Goudfrooij, P.  2000,
   AJ, 120, 801
%
\bibitem[Ribas et~al.(2005)]{Ribas:etal:05}
   Ribas, I., Jordi, C., Vilardell, F., Fitzpatrick, E.~L., 
   Hilditch, R.~W., \& Guinan, E.~F. 2005,
   ApJ, 635, 37
%
\bibitem[Riess et~al.(2005)]{Riess:etal:05} 
   Riess, A.~G., et~al. 2005,
   ApJ, 627, 579
%
\bibitem[Rizzi et~al.(2007)]{Rizzi:etal:07}
   Rizzi, L., Tully, R.~B., Makarov, D., Makarova, L., Dolphin, A.~E.,
   Sakai, S., \& Shaya, E.~J. 2007, 
   ApJ, 661, 815   
%
\bibitem[Romaniello et~al.(2005)]{Romaniello:etal:05}
   Romaniello, M., Primas, F., Mottini, M., Groenewegen, M., Bono, G.,
   \& Francois, P. 2005, 
   A\&A, 429, L37
%
\bibitem[Rood(1972)]{Rood:72}
   Rood, R.~T. 1972, 
   ApJ, 177, 681
%
\bibitem[Russell(1913)]{Russell:13}
   Russell, H.~N. 1913, 
   Science, 37, 651
%
\bibitem[Saha \& Hoessel(1990)]{Saha:Hoessel:90}
   Saha, A., \& Hoessel, J.~G. 1990,
   AJ, 99, 97   (NGC\,185)
%
\bibitem[Saha et~al.(1992)]{Saha:etal:92}
   Saha, A., Hoessel, J.~G., \& Krist, J. 1992,
   AJ, 103, 84   (NGC\,205)
%
\bibitem[Saha et~al.(1990)]{Saha:etal:90}
   Saha, A., Hoessel, J.~G., \& Mossman, A.~E. 1990,
   AJ, 100, 108   (NGC\,147)
%
\bibitem[Saha et~al.(1986)]{Saha:etal:86}
   Saha, A., Monet, D.~G., \& Seitzer, P. 1986,
   AJ, 92, 302   (Carina)
%
\bibitem[STT\,06(2006)Saha et~al.]{STT:06}
   Saha, A., Thim, F., Tammann, G.~A., Reindl, B., \& Sandage, A. 2006,
   ApJS, 165, 108   (STT\,06)
%
\bibitem[Sakai et~al.(2004)]{Sakai:etal:04}
   Sakai, S., Ferrarese, L., Kennicutt, R.~C., \& Saha, A. 2004,
   ApJ, 608, 42
%
\bibitem[Sakai et~al.(1997)]{Sakai:etal:97}
   Sakai, S., Madore, B.~F., Freedman, W.~L., Lauer, T.~R., 
   Ajhar, E.~A., \&  Baum, W.~A. 1997,
   ApJ, 478, 49
%
\bibitem[Salaris \& Cassisi(1997)]{Salaris:Cassisi:97}
   Salaris, M., \& Cassisi, S. 1997, 
   MNRAS, 289, 406
%
\bibitem[Salaris \& Cassisi(1998)]{Salaris:Cassisi:98}
   Salaris, M., \& Cassisi, S. 1998, 
   MNRAS, 298, 166
%
\bibitem[Salaris et~al.(1997)]{Salaris:etal:97}
   Salaris, M., degl'Innocenti, S., \& Weiss, A. 1997, 
   ApJ, 479, 665
%
\bibitem[Salaris et~al.(2003)]{Salaris:etal:03}
   Salaris, M., Percival, S., \& Girardi, L. 2003,
   MNRAS, 345, 1030
%
\bibitem[Sandage(1971)]{Sandage:71}
   Sandage, A. 1971,
   in Pont. Acad. Scient. Scripta Varia No. 35, Nuclei of Galaxies,
   ed. D.~J.~K. O'Connell,
   (Amsterdam: North Holland), 601
%
\bibitem[Sandage(1988)]{Sandage:88}
   Sandage, A. 1988,
   PASP, 100, 935
%
\bibitem[Sandage(2006)]{Sandage:06}
   Sandage, A., 2006, 
   AJ, 131, 1750
%
\bibitem[Sandage et~al.(1999)]{Sandage:etal:99}
   Sandage, A., Bell, R.~A., \& Tripicco, M.~J. 1999,
   ApJ, 522, 250
%
\bibitem[ST\,06(2006)Sandage \& Tammann]{Sandage:Tammann:06}
   Sandage, A., \& Tammann, G.~A. 2006,
   ARA\&A, 44, 93 (ST\,06)
%
\bibitem[STR\,04(2004)Sandage et~al.]{STR:04}
   Sandage, A., Tammann, G.~A., \& Reindl, B. 2004,
   A\&A, 424, 43  (STR\,04)
%
\bibitem[STS\,06(2006)Sandage et~al.]{STS:06}
   Sandage, A., Tammann, G.~A., Saha, A., Reindl, B., Macchetto,
   F.~D., \& Panagia, N. 2006,
   ApJ, 653, 843   (STS\,06)
%
\bibitem[Sarajedini et~al.(2006)]{Sarajedini:etal:06}
   Sarajedini, A., Barker, M.~K., Geisler, D., Harding, P., \& 
   Schommer, R. 2006, 
   AJ, 132, 1361   (NGC\,598)
%
\bibitem[Schlegel et~al.(1998)]{Schlegel:etal:98}
   Schlegel, D., Finkbeiner, D., \& Davis, M. 1998,
   ApJ, 500, 525
%
\bibitem[Seares et~al.(1930)]{Seares:etal:30}
   Seares, F.~H., Kapteyn, J.~C., \& van Rhijn P.~J. 1930, 
   Mount Wilson Catalogue of Photographic Magnitudes in Selected Areas 1-139,
   Carnegie Institution of Washington Pub. 402
%
\bibitem[Seth et~al.(2005)]{Seth:etal:05}
   Seth, A.~C., Dalcanton, J.~J., \& de~Jong, R.~S. 2005,
   AJ, 129, 1331
%
\bibitem[Shapley(1918)]{Shapley:18}
   Shapley, H. 1918, 
   ApJ, 48, 89 (Paper VI of his series)
%
\bibitem[Siegel \& Majewski(2000)]{Siegel:Majewski:00}
   Siegel, M.~H., \& Majewski, S.~R. 2000,
   AJ, 120, 284   (Leo~II)
%
\bibitem[Smecker-Hane et~al.(1994)]{Smecker-Hane:etal:94}
   Smecker-Hane, T.~A., Stetson, P.~B., Hesser, J.~E., \& Lehnert, M.~D. 1994,
   AJ, 108, 507
%
\bibitem[Smith et~al.(1992)]{Smith:etal:92}
   Smith, H.~A., Silbermann, N.~A., Baird, S.~R., \& Graham, J.~A. 1992, 
   AJ, 104, 1430
%
\bibitem[Sollima et~al.(2006)]{Sollima:etal:06}
   Sollima, A., Cacciari, C., \& Valenti, E. 2006,
   MNRAS, 372, 1675
%
\bibitem[Soszynski et~al.(2006)]{Soszynski:etal:06}
   Soszynski, I., Gieren, W., Pietrzynski, G., Bresolin, F.,
   Kudritzki, R.-P., \& Storm, J. 2006, 
   ApJ, 648, 375 (NGC\,3109)
%
\bibitem[Soszynski et~al.(2002)]{Soszynski:etal:02}
   Soszynski, I., et~al. 2002, 
   AcA, 52, 369   (SMC)
%
\bibitem[Soszynski et~al.(2003)]{Soszynski:etal:03}
   Soszynski, I., et~al. 2003, 
   AcA, 53, 93   (LMC)
%
\bibitem[Spergel et~al.(2007)]{Spergel:etal:07}
   Spergel, D.~N., et~al. 2007, 
   ApJS, 170, 377
%
\bibitem[Storm et~al.(2004)]{Storm:etal:04}
   Storm, J., Carney, B.~W., Gieren, W.~P., Fouqu{\'e}, P., Latham,
   D.~W., \& Fry, A.~M. 2004,
   A\&A, 415, 531
%
\bibitem[Sweigart \& Gross(1978)]{Sweigart:Gross:78}
   Sweigart, A.~V., \& Gross, R.~G. 1978,
   ApJS, 36, 405
%
\bibitem[Tammann \& Reindl(2002)]{Tammann:Reindl:02}
   Tammann, G.~A., \&  Reindl, B. 2002,
   Ap \& Space Sci., 280, 165
%
\bibitem[Tammann et~al.(2002)]{Tammann:etal:02}
   Tammann, G.~A., Reindl, B., Thim, F., Saha, A., \& Sandage, A. 2002,
   in A New Era in Cosmology,
   eds. T.~Shanks, \& N.~Metcalfe
   (San Francisco: ASP), 258
%
\bibitem[TSR\,03(2003)Tammann et~al.]{TSR:03}
   Tammann, G.~A., Sandage, A., \& Reindl, B. 2003, 
   A\&A, 404, 423 (TSR\,03)
%
\bibitem[Tanvir et~al.(2005)]{Tanvir:etal:05}
   Tanvir, N.~R., Hendry, M.~A., Watkins, A., Kanbur, S.~M.,
   Berdnikov, L.~N., \& Ngeow, C.~C. 2005, 
   MNRAS, 363, 749
%
\bibitem[Tegmark et~al.(2006)]{Tegmark:etal:06}
   Tegmark, M., et~al. 2006, 
   Phys. Rev. D, 74, 123507 
%
\bibitem[Thackery(1954)]{Thackery:54}
   Thackery, A.~D. 1954, 
   %
   in Trans. IAU, VIII (Rome 1952 meeting),
   Report of Commission 28, (Cambridge: Cambridge Univ. Press), 397
%
\bibitem[Thackery(1958)]{Thackery:58}
   Thackery, A.~D. 1958, 
   %
   report to the 1957 Vatican Conference on Stellar Populations, 
   Specola Vaticana, Vol. 5, ed. D.~J.~K. O'Connell, 195
%
\bibitem[Thim et~al.(2003)]{Thim:etal:03}
   Thim, F., Tammann, G.~A., Saha, A., Dolphin, A., Sandage, A.,
   Tolstoy, E., \& Labhardt, L. 2003,
   ApJ, 590, 256
%
\bibitem[Thuan \& Gunn(1976)]{Thuan:Gunn:76}
   Thuan, T.~X., \& Gunn, J.~E. 1976, 
   PASP, 88, 543
%
\bibitem[Tully \& Pierce(2000)]{Tully:Pierce:00}
   Tully, R.~B., \& Pierce, M.~J. 2000, 
   ApJ, 533, 744
%
\bibitem[Tully et~al.(2006)]{Tully:etal:06}
   Tully, R.~B., et~al. 2006,
   AJ, 132, 729 
%
\bibitem[Udalski(1998)]{Udalski:98}
   Udalski, A. 1998, 
   AcA, 48, 113
%
\bibitem[Udalski(2000)]{Udalski:00}
   Udalski, A. 2000, 
   AcA, 50, 279 (Carina)
%
\bibitem[Udalski et~al.(1999a)]{Udalski:etal:99a}
   Udalski, A., Soszynski, I., Szymanski, M., Kubiak, M., Pietrzynski,
   G., Wozniak, P., \& Zebrun, K. 1999a, 
   AcA, 49, 223
%
\bibitem[Udalski et~al.(1999b)]{Udalski:etal:99b}
   Udalski, A., Soszynski, I., Szymanski, M., Kubiak, M., Pietrzynski,
   G., Wozniak, P., \& Zebrun, K. 1999b, 
   AcA, 49, 437
%
\bibitem[Udalski et~al.(1999c)Udalski's et~al.]{Udalski:etal:99c}
   Udalski, A., Szymanski, M., Kubiak, M., Pietrzynski, G., Szymanski,
   M., Wozniak, P., \& Zebrun, K. 1999c, 
   AcA, 49, 201
%
\bibitem[VandenBerg et~al.(2000)]{VandenBerg:etal:00}
   VandenBerg, D.~A., Swenson, E.~J., Rogers, F.~J., Iglesias,
   C.~A., \& Alxander, D.~R. 2000, 
   ApJ, 532, 430
%
\bibitem[van den Bergh(1995)]{vandenBergh:95}
   van den Bergh, S. 1995, 
   ApJ, 446, 39
%
\bibitem[van Leeuwen et~al.(2007)]{vanLeeuwen:etal:07}
   van Leeuwen, F., Feast, M.~W., Whitelock, P.~A., \& Laney,
   C.~D. 2007,  
   MNRAS, 379, 723   
%
\bibitem[Vilardell et~al.(2007)]{Vilardell:etal:07}
   Vilardell, F., Jordi, C., \& Ribas, I. 2007,  
   A\&A, 473, 847
%
\bibitem[Walker \& Mack(1988)]{Walker:Mack:88}
   Walker, A., \& Mack, A.~R. 1988, 
   AJ, 96, 872
%
\bibitem[Yahil et~al.(1980)]{Yahil:etal:80} 
   Yahil, A., Sandage, A., \& Tammann, G.~A. 1980, 
   in Physical Cosmology, eds. E.~Balian, J.~Audouze, \& D.~N.~Schramm
   (Amsterdam: North-Holland), 127
%
\bibitem[Yahil et~al.(1977)]{Yahil:etal:77} 
   Yahil, A., Tammann, G.~A., \& Sandage, A. 1977, 
   ApJ, 217, 903
%
\bibitem[Zaritsky et~al.(1994)]{Zaritsky:etal:94}
   Zaritsky, D., Kennicutt, R.~C., \& Huchra, J.~P. 1994, 
   ApJ, 420, 87
%
\bibitem[Zinn \& West(1984)]{Zinn:West:84}
   Zinn, R., \& West, M.~J. 1984, 
   ApJS, 55, 45
%
\bibitem[Zoccali et~al.(2003)]{Zoccali:etal:03}
   Zoccali, M., et~al. 2003, 
   A\&A, 399, 931
\end{thebibliography}
\end{document}